\documentclass[aps,prb,tightenlines,showpacs,twocolumn,superscriptaddress,10pt]{revtex4-1}  

\usepackage{amsmath,amssymb,amsfonts,bm}
\usepackage{graphicx}
\usepackage{dcolumn}
\usepackage{color}
\usepackage[colorlinks=true,linkcolor=blue,citecolor=blue]{hyperref}%

\begin{document}


\title{Resonance states near a quantum magnetic impurity in single-layer FeSe superconductors with $d$-wave symmetry}
\author{Liang Chen}
\email[Corresponding Email: ]{slchern@ncepu.edu.cn}
\affiliation{School of Mathematics and Physics, North China Electric Power University, Beijing, 102206, China}
\author{Wei-Wei Zhao}
\affiliation{School of Mathematics and Physics, North China Electric Power University, Beijing, 102206, China}
\author{Rong-Sheng Han}
\affiliation{School of Mathematics and Physics, North China Electric Power University, Beijing, 102206, China}
\date{\today}

\begin{abstract}
In this work, we investigate the local density of states (LDOS) near a magnetic impurity in single-layer FeSe superconductors. The two-orbital model with spin-orbit coupling proposed in Ref. [{\emph{Phys. Rev. Lett.} \textbf{119}, 267001 (2017)}] is used to describe the FeSe superconductor. In the strong coupling regime, two impurity resonance peaks appear with opposite resonance energies in the LDOS spectral function. For a strong spin-orbit coupling, the superconducting gap in this model is $d$-wave symmetric with nodes, the spatial distributions of the LDOS at the two resonance energies are fourfold symmetric, which reveals typical characteristic of $d$-wave pairing. When the spin-orbit coupling is not strong enough to close the superconducting gap, we find that the spatial distribution of the LDOS at one of the resonance energies manifests $s$-wave symmetry, while the pairing potential preserves $d$-wave symmetry. This result is consistent with previous experimental investigations. 
\end{abstract}


\maketitle


\section{Introduction} \label{sec1} 
The magnetic scattering problem in normal metals \cite{Hewsonbook} and superconductors \cite{BalatskyAV2006RMP} plays an important role in understanding the quantum state of hosting materials. The interaction between magnetic impurity and conducting electron states in normal metals leads to the famous Kondo effects \cite{Hewsonbook}. In superconductors, the magnetic impurity induces the localized  Yu-Shiba-Rusinov quasiparticle states \cite{Yu1965APS,Shiba1968CSS,Rusinov1969JETPL}. Like the standard phase-sensitive tetracrystal measurements \cite{Tsuei1994PRL,Tsuei1997Nature} and quasiparticle interference experiments \cite{Hoffman2002Science,WangQH2003PRB,Hanaguri2007NatPhys,Hanaguri2009Science}, the LDOS of the resonance states near the magnetic (and nonmagnetic) impurity is an important method to uncover the symmetries of pairing potentials in unconventional superconductors, i.e., the high-temperature cuprates \cite{PanSH2000Nat,HudsonEW2001Nat,BobroffJ2001PRL,ZhangGM2001PRL,Polkovnikov2001PRL,ZhuJX2000PRB,VojtaM2002PRB,Polkovnikov2002PRB,DaiX2003PRB,BaarS2016JSNM}, iron-based superconductors \cite{TsaiWF2009PRB,BangY2009PRB,AkbariA2010PRB}, chiral $p$-wave superconductors and topological superconductors \cite{SauJD2013PRB,FuZG2012JPCM,ZhaGQ2017EPL,GuoYW2017FP}, etc. 

As one of the simplest iron-based high-temperature superconductor, the iron-chalcogenide compound FeSe has a transition temperature $T_c\approx9$K at ambient pressure \cite{HsuFC2008PNAS}. The Fermi surface of FeSe bulk material displays a hole pocket near the $\Gamma$-point and two electron pockets around the $M$-point in the Brillouin zone. The neutron scattering measurements \cite{WangQNatMat2015} show that the superconductivity is driven by stripe antiferromagnetic fluctuations and a sign-changing pairing symmetry (i.e., $s_{\pm}$-pairing) is more likely.  The coexistence of hole pocket and electron pockets in the Brillouin zone is essential for the stripe antiferromagnetic fluctuations.
Recently, the single-layer FeSe films grown on Nb-doped SrTiO$_3$ substrate have attracted much attention due to the remarkable high critical temperature $T_c\sim100$K \cite{WangQY2012CPL,Liu2012NatComm,HeSL2013NatMat,TanSY2013NatMat,LeeJJ2014NatLett,GeJF2014NatMat}. Angle-resolved photoemission spectroscopy (ARPES) measurements show that \cite{Liu2012NatComm} the energies of fermions near the $\Gamma$-point are about 80meV lower than the Fermi energy, which demonstrates that the hole pocket of the Fermi surface near the $\Gamma$-point is excluded. Various possible pairing symmetries have been proposed for such systems with only the electron pockets, e.g., plain $s$-wave pairing \cite{FangC2011PRX,ZhouY2011EPL,YangF2013PRB}, nodeless $d$-wave pairing \cite{MaierTA2011PRB,WangF2011EPL}, new extended $s$-wave pairing \cite{MazinII2011PRB}, etc. The quasiparticle interference experiment and scanning tunnelling microscopy (STM) topograph of resonance states near magnetic and nonmagnetic impurities suggest that the pairing potential should be plain $s$-wave \cite{FanQ2015NatPhys}. 

In addition to the non-existence of hole pocket near the $\Gamma$-point, the single-layer FeSe on SrTiO$_3$ has another property being different from the bulk material, the inversion-symmetry is breaking and the electric field at the interference will induce a spin-orbit coupling. Kang and Fernandes have studied the superconductivity induced by nematic fluctuations in single-layer FeSe \cite{KangJ2016PRL}. They find that the spin-orbit coupling and inversion-symmetry breaking play important roles in lifting the degeneracy of $s$-wave and $d$-wave pairing symmetries, and helping to select the $s$-wave state which is agreement with experimental results shown in Refs. [\onlinecite{FanQ2015NatPhys}] and [\onlinecite{ZhangY2016PRL}]. Recently, Agterberg et. al. find another explanation \cite{AgterbergDF2017PRL} for the superconductivity in single-layer FeSe using a spin-fermion model coupled to fluctuations of checkerboard magnetic order. A fully gapped, nodeless, $d$-wave pairing potential is predicted in this theory if the energy scale of the relevant spin-orbit coupling is smaller than the superconducting gap. The ARPES results in Ref. [\onlinecite{ZhangY2016PRL}] are well explained by this model using proper relevant parameters. A nature question is, is this $d$-wave symmetric nodeless superconductivity proposed in Ref. [\onlinecite{AgterbergDF2017PRL}] applicable for the explanation of the $s$-wave symmetric resonance states observed by STM \cite{FanQ2015NatPhys}. As far as we know, both the resonance states near impurities and quasiparticle interference patterns shown in Ref. [\onlinecite{FanQ2015NatPhys}] are not studied for the nodeless $d$-wave pairing potential. 

In this work, we study the LDOS of the resonance states near a magnetic impurity in single-layer FeSe superconductors with $d$-wave pairing potential. The paper is organized as follows: we propose the theoretical model in Sec. \ref{sec2}, study the LDOS for different parameters in Sec. \ref{sec3}, and give a conclusion in Sec. \ref{sec4}. 

\section{Theoretical Model}\label{sec2} 

\begin{figure}[t]
	\includegraphics[width=\columnwidth]{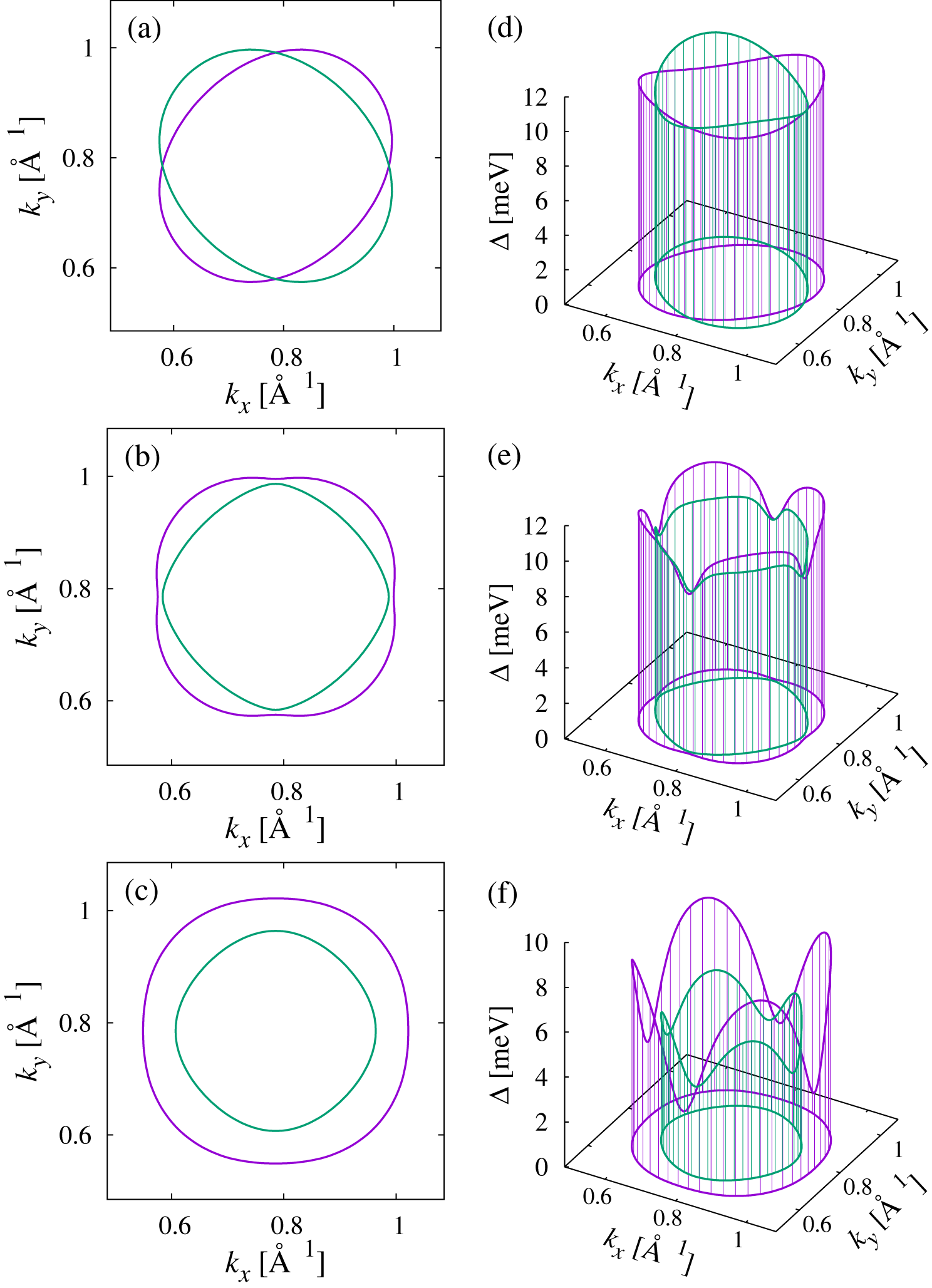} 
	\caption{(color online) The Fermi surfaces and the anisotropic superconducting gap on the Fermi surfaces for different spin-orbit couplings. (a) and (d): $v_{\rm{so}}=0$, (b) and (e): $v_{\rm{so}}=12$meV{\AA},  (c) and (f): $v_{\rm{so}}=80$meV{\AA}.}%
	\label{fig1}%
\end{figure}

The total Hamiltonian used to model the magnetic impurity in single-layer FeSe is consist of three parts, 
\begin{equation}
	H=H_{\rm{imp}}+H_{\rm{hyb}}+H_{\rm{sc}}, \label{eq1}
\end{equation}
where $H_{\rm{imp}}$ describes the impurity located at $\bm{R}=0$ with on-site Coulomb interaction, 
\begin{equation}
	H_{\rm{imp}}=\sum_{s=\uparrow,\downarrow}\varepsilon_{d}d_{s}^{\dagger}d_s+Un_{\uparrow}n_{\downarrow}, \label{eq2}
\end{equation}
$\varepsilon_{d}$ is the impurity energy, $s=\uparrow,\downarrow$ is the spin of the electron state on the impurity, $d_{s}^{\dagger}$ and $d_s$ are the creation and annihilation operators of the impurity states with spin-$s$, $U$ refers to the on-site Coulomb interaction, $n_{s}=d_{s}^{\dagger}d_s$ is the occupation number operator of spin-$s$ state. 
\begin{equation}
	H_{\rm{hyb}}=\frac{1}{\sqrt{N}}\sum_{\bm{k},s}V_{\bm{k}}\left(c_{\bm{k},s}^{\dagger}d_{s}+d_s^{\dag}c_{\bm{k},s}\right), \label{eq3}  
\end{equation}
is the hybridization between the magnetic impurity and the superconducting states, where $V_{\bm{k}}$ is the hybridization energy between the impurity state and the superconducting state with wave-vector $\bm{k}=(k_x,k_y)$. Hereafter we set the hybridization to be short-ranged, so that it is wave-vector independent, $V_{\bm{k}}=V_0$. $c_{\bm{k},s}^{\dagger}$ and $c_{\bm{k},s}$ are the creation and annihilation operators of the electron state in single-layer FeSe with wave-vector $\bm{k}$ and spin-$s$. $N$ refers to the total number of wave-vectors in summation.  The last term in Eq. (\ref{eq1}) describes the free Hamiltonian of the superconducting state \cite{AgterbergDF2017PRL}, 
\begin{gather}
	H_{\rm{sc}}=\sum_{\bm{k}}\psi_{\bm{k}}^{\dag}\left\{\varepsilon_0(\bm{k})\tau_0\sigma_0+\gamma_{xy}(\bm{k})\tau_z\sigma_0\right.\notag\\
	\left.+\tau_x\left[\gamma_y(\bm{k})\sigma_x+\gamma_x(\bm{k})\sigma_y\right]\right\}\psi_{\bm{k}}\notag\\
	+\frac{1}{2}\sum_{\bm{k}}\left[\psi_{-\bm{k}}^{T}(\Delta_d\tau_0+\Delta_z\tau_z)i\sigma_y\psi_{\bm{k}}+h.c.\right]. \label{eq4}
\end{gather}
Here $\psi_{\bm{k}}=(c_{\bm{k},1,\uparrow},c_{\bm{k},2,\uparrow},c_{\bm{k},1,\downarrow},c_{\bm{k},2,\downarrow})^{\mathsf{T}}$ (the superscript $\mathsf{T}$ refers to matrix transpose) is the four-component spinor description of the electron states with two orbital degrees of freedom described by $\tau_{x,y,z}$ Pauli matrices and two spin degrees of freedom described by $\sigma_{x,y,z}$ Pauli matrices. $\tau_0$ and $\sigma_0$ are the $2\times2$ identity matrices in orbit and spin space, respectively. $h.c.$ means the Hermitian conjugate.  $\varepsilon_0(\bm{k})\pm\gamma_{xy}(\bm{k})$ are the dispersion of the two orbits. They are given in the following tight-binding form \cite{Korshunov2008PRB},  
\begin{gather}
	\varepsilon_0(\bm{k})=t_1\left[\cos(k_xa)+\cos(k_ya)\right]-\epsilon, \label{eq5}\\
	\gamma_{xy}(\bm{k})=t_2\cos\left(\frac{k_xa}{2}\right)\cos\left(\frac{k_ya}{2}\right). \label{eq6}
\end{gather}
$\gamma_x(\bm{k})$ and $\gamma_y(\bm{k})$ in Eq. (\ref{eq4}) represent the spin-orbit couplings, they are given by, 
\begin{equation}
	\gamma_x(\bm{k})=-t_3\sin(k_xa), \hspace{3pt}\gamma_y(\bm{k})=-t_3\sin(k_ya). \label{eq7}
\end{equation}
The superconducting gap terms are given by $\Delta_d=\Delta_2\sin(k_xa)\sin(k_ya)$ and $\Delta_z=\Delta_0$ with $\Delta_0=11$meV and $\Delta_2=-2.34$meV. 
$a$ in Eqs. (\ref{eq5})-(\ref{eq7}) is the lattice constant, it is set to be $4${\AA}  in this work \cite{TanSY2013NatMat,CaoHY2014PRB}. The other parameters are chosen as follows, $t_1=171.875$meV, $t_2=150$meV, $\epsilon=-288.75$meV, $t_3=v_{\rm{so}}/a$, and $v_{\rm{so}}$ represents the spin-orbit coupling strength. One can check that, in the continuous limit, the model Hamiltonian (\ref{eq4}) tends to that given in Ref. [\onlinecite{AgterbergDF2017PRL}]. Fig. \ref{fig1} shows the Fermi surfaces and superconducting gap on the Fermi surfaces for different spin-orbit couplings, one can find that they are consistent with those given in Ref. [\onlinecite{AgterbergDF2017PRL}]. 

Now we study the model Hamiltonian (\ref{eq1}). In the strong Coulomb interaction limit, $U\rightarrow\infty$, the double occupied state of electrons on the impurity can be excluded. This limit may be represented by introducing an auxiliary boson operator $b$ to reformulate the creation and annihilation operators of the impurity states, $(d_s^{\dag},d_s)=(f_s^{\dag}b,b^{\dag}f_s)$. The extra degrees of freedom after introducing these boson operators are restricted by the constraint $\hat{Q}=b^{\dag}b+\sum_{s}f_s^{\dag}f_s=1$. In the mean-field approximation, $b$ and $b^{\dag}$ are replaced by their expectation value, $b_0=\langle{b}\rangle=\langle{b^{\dag}}\rangle$, and the constraint is approximated by adding a term $\lambda_0\left(b^{\dag}b+\sum_{s}f_s^{\dag}f_s\right)$ to the Hamiltonian, where $\lambda_0$ is a Lagrangian multiplier, it renormalizes the impurity energy. Both $b_0$ and $\lambda_0$ need to be determined self-consistently by minimizing the free energy. The mean-field Hamiltonian is given by, 
\begin{align}
	H_{\rm{MF}}=\tilde{H}_{\rm{imp}}+\tilde{H}_{\rm{hyb}}+H_{\rm{sc}}+\lambda_0\left(b_0^2-1\right), \\
	\tilde{H}_{\rm{imp}}=\sum_{s}\tilde{\varepsilon}_df_s^{\dag}f_s, \hspace{5pt}\tilde{H}_{\rm{hyb}}=\frac{\tilde{V}_0}{\sqrt{N}}\sum_{\bm{k},s}c_{\bm{k},s}^{\dagger}f_{s}+h.c.,
\end{align}
where $\tilde{\varepsilon}_d=\varepsilon_d+\lambda_0$ is the renormalized impurity energy, $\tilde{V}_0=b_0V_0$ is the renormalized hybridization. In the Bogoliubov-deGennes (BdG) formalism, the mean-field Hamiltonian can be recast as, 
\begin{align}
	\mathcal{H}_{\rm{MF}}&=\lambda_0(b_0^2-1)+\tilde{\varepsilon}_d+\frac{1}{2}\Phi^{\dag}\Lambda\Phi+\frac{1}{2}\sum_{\bm{k}}\Psi_{\bm{k}}^{\dag}h_{\rm{BdG}}\Psi_{\bm{k}}\notag \\
	&+\frac{1}{2\sqrt{N}}\sum_{\bm{k}}\left(\Psi_{\bm{k}}^{\dag}\bar{V}\Psi+\Psi^{\dag}\bar{V}^{\dag}\Psi_{\bm{k}}\right),
\end{align}
where $\Phi=(f_{\uparrow},f_{\downarrow},f_{\uparrow}^{\dag},f_{\downarrow}^{\dag})^{\mathsf{T}}$ and $\Psi_{\bm{k}}=\left[\psi_{\bm{k}},\left(\psi_{-\bm{k}}^{\dag}\right)^{\mathsf{T}}\right]^{\mathsf{T}}$ are the Nambu spinors. The matrices $\Lambda$ and $h_{BdG}$ are given by
\begin{align}
	&\Lambda=\tilde{\varepsilon}_d\varsigma_z\tau_0\sigma_0, \\
	&h_{\rm{BdG}}=\varepsilon_0(\bm{k})\varsigma_z\tau_0\sigma_0+\gamma_{xy}(\bm{k})\varsigma_z\tau_z\sigma_0+\gamma_{y}(\bm{k})\varsigma_0\tau_x\sigma_x \notag \\
	&+\gamma_{x}(\bm{k})\varsigma_z\tau_x\sigma_y-\varsigma_y\left(\Delta_d\tau_0+\Delta_z\tau_z\right)\sigma_y, 
\end{align}
where $\varsigma_0$ and $\varsigma_{x,y,z}$ are the identity matrix and Pauli matrices in the Nambu spinor space, respectively. $\bar{V}$ is a $8\times4$ matrix representation of the renormalized hybridization, whose elements are given by, $\bar{V}_{ij}=\tilde{V}_0(\delta_{i,2j-1}+\delta_{i,2j})$. 

\begin{figure}[tb]
	\includegraphics[width=\columnwidth]{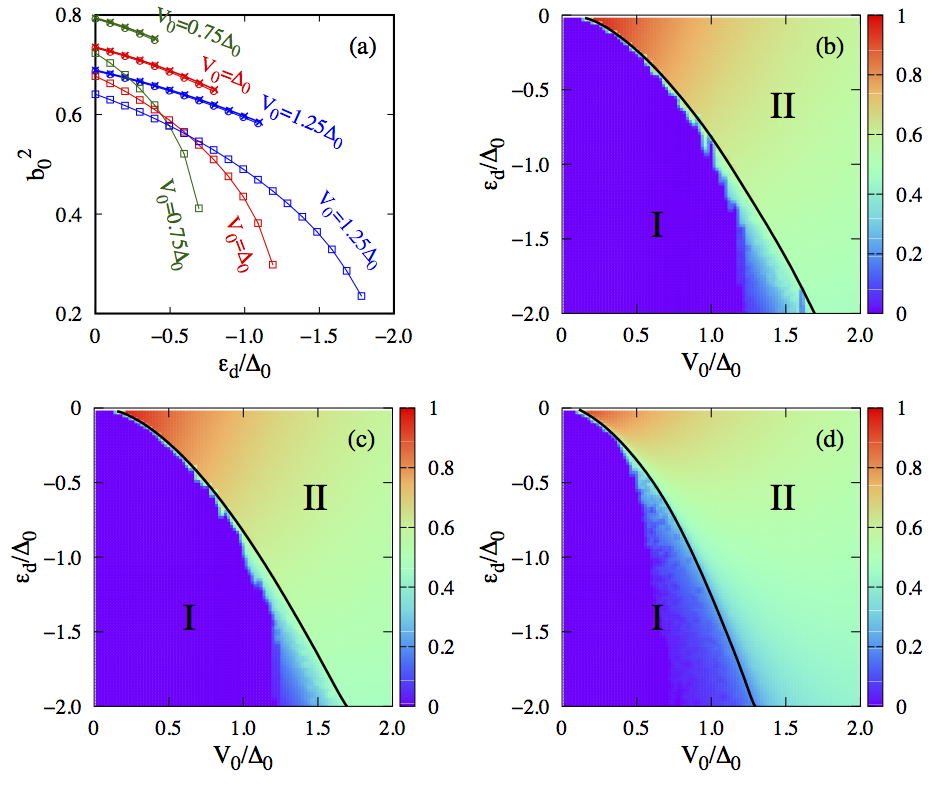} 
	\caption{(color online) (a) The numerical solutions of $b_0^2$ as functions of $\varepsilon_d/\Delta_0$ for different hybridizations and different spin-orbit couplings. The blue lines, red lines and green lines show the results for hybridizations $V_0=1.25\Delta_0$, $\Delta_0$ and $0.75\Delta_0$, respectively. The lines labeled with crosses ($\times$), circles ($\circ$) and squares ($\scriptstyle\Box$) represent the spin-orbit coupling $v_{\rm{so}}=0$, $12$meV{\AA} and $80$meV{\AA}, respectively. The results for $v_{\rm{so}}=0$ and $v_{\rm{so}}=12$meV{\AA} are very close to each other. (b)-(d) show the contour plots of $b_0^2$ vs. hybridization $V_0/\Delta_0$ and impurity energy $\varepsilon_{d}/\Delta_0$ for the three different spin-orbit coupling strengths, $v_{\rm{so}}=0$, $12$meV{\AA} and $80$meV{\AA}. The black lines in (b)-(d) show the boundaries between the isolated magnetic moment regime (I) and the strong coupling regime (II). }%
	\label{fig2}%
\end{figure}

\begin{figure*}[t]
	\includegraphics[width=\textwidth]{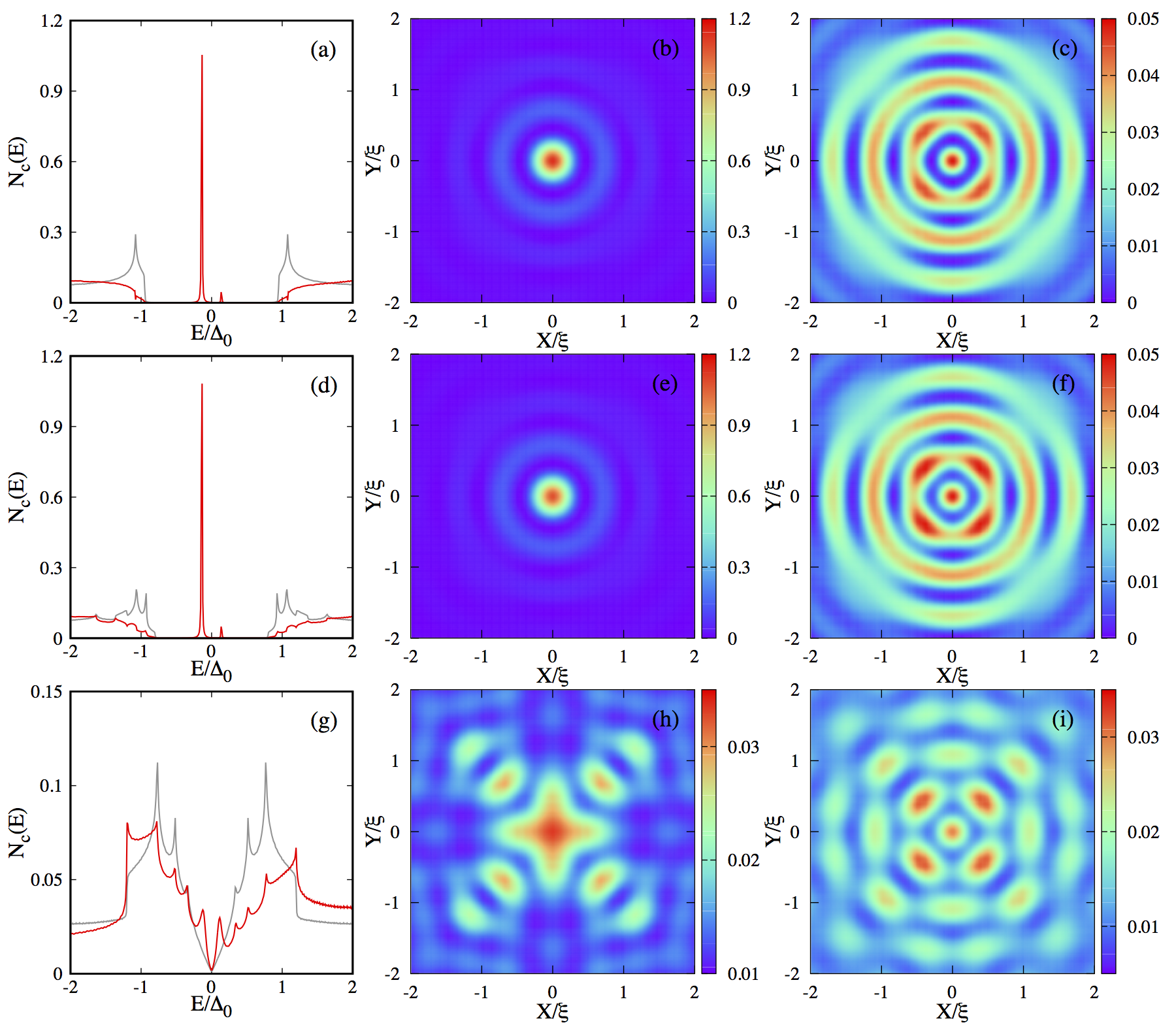} 
	\caption{(color online) The first column: LDOS of the conduction electrons for $v_{\rm{so}}=0$, $12$meV{\AA} and $80$meV{\AA} from top to bottom. The gray lines show the LDOS for single-layer FeSe superconducting states without impurities. The red lines present the LDOS close to the impurity ($\bm{R}=0$). The second column: spatial distribution of LDOS at $E=\Omega_{r}$, where $\Omega_{r}<0$ is the position of the first resonance peak. The third column: spatial distruibution of LDOS at the second resonance peak, $E=-\Omega_r$. The parameters for numerical calculations are chosen as follows: $\varepsilon_{d}=-\Delta_0/2$, $V_0=\Delta_0$, and $v_{\rm{so}}$s take the same value for each row. The intensity of LDOS is meV$^{-1}$, the coherent length $\xi=\hbar{v_F}/\Delta_0$ is about 25{\AA}, or roughly six lattice constant in single-layer FeSe. The maximum intensities in (b), (c), (e), (f), and (h) are located at $\bm{R}=0$, in figure (i), it is located at about $\bm{R}=(\pm0.5\xi,\pm0.5\xi)$. }%
	\label{fig3}%
\end{figure*}

Using the standard functional integral techniques, we find that the free energy is given by, 
\begin{equation}
	\mathcal{F}=\lambda_0\left(b_0^2-1\right)+\tilde{\varepsilon}_d+\frac{1}{2}\frac{1}{\beta}\sum_{n}{\rm{tr}}\ln\left[G_{f}(i\omega_n)\right], \label{eq13}
\end{equation}
where $\beta=1/k_BT$ is the inverse temperature, $k_B$ is the Boltzmann constant and $T$ represents temperature, $\omega_n=\pi(2n+1)/\beta$ is the Matsubara frequency, $G_{f}(i\omega_n)=\left[i\omega_n-\Lambda-\Sigma_{f}(i\omega_n)\right]^{-1}$ is the Green's function of the impurity states expressed in imaginary-frequency representation, $\Sigma_{f}(i\omega_n)=\frac{1}{N}\sum_{\bm{k}}\bar{V}^{\dag}G_{c}^{(0)}(i\omega_n,\bm{k})\bar{V}$ is the self-energy, and $G_{c}^{(0)}(i\omega_n,\bm{k})=(i\omega_n-h_{\rm{BdG}})^{-1}$ is the unperturbed Green's function of the electrons in single-layer FeSe. Minimizing the free energy, Eq. (\ref{eq13}), we can find the self-consistent equations of $b_0$ and $\lambda_0$, 
\begin{gather}
	b_0^2+\frac{1}{2\beta}\sum_{n}{\rm{tr}}\left[G_{f}(i\omega_n)\varsigma_z\sigma_0\right]=0, \label{eq14} \\
	\lambda_0b_0^2+\frac{1}{2\beta}\sum_{n}{\rm{tr}}\left[G_{f}(i\omega_n)\Sigma_{f}(i\omega_n)\right]=0. \label{eq15}
\end{gather}
The LDOS near the magnetic impurity is obtained by using the analytic continuation, $i\omega_n\rightarrow{E}+i0^+$, according to 
\begin{equation}
	N_{c}(E,\bm{R})=-\frac{1}{\pi}{\rm{Im}}\left[G_c(E;\bm{R},\bm{R})\frac{1+\varsigma_z\tau_0\sigma_0}{2}\right],
\end{equation}
where the Green's function of the conduction electrons are determined by, 
\begin{gather}
	G_c(E;\bm{R},\bm{R})=\frac{1}{N}\sum_{\bm{k},\bm{k}'}e^{i(\bm{k}-\bm{k}')\cdot\bm{R}}G_c(E;\bm{k},\bm{k}'), \\
	G_c(E;\bm{k},\bm{k}')=G_c^{(0)}(E,\bm{k})\left[\delta_{\bm{k},\bm{k}'}+T(E)G_c^{(0)}(E,\bm{k}')\right].
\end{gather}
Here $T(E)=\bar{V}G_{f}(E)\bar{V}^{\dag}/N$ is the $T$-matrix, $G_c^{(0)}(E,\bm{k})$ is the analytic continuation of $G_c^{(0)}(i\omega_n,\bm{k})$. By solving Eqs. (\ref{eq14}) and (\ref{eq15}), we can get the values of $\tilde{\varepsilon}_d$ and $\tilde{V}_0$, the Green's functions $G_{f}(E)$ and $G_c(E;\bm{k},\bm{k}')$, and the LDOS $N_c(E,\bm{R})$.

\section{Numerical results}\label{sec3}
It is difficult to find the analytic solutions of $\lambda_0$ and $b_0$ due to the complex band structure. Here we show the numerical results. Fig. \ref{fig2}(a) shows $b_0^2$ vs. $\varepsilon_{d}$ for different hybridizations and different spin-orbit couplings. One can find that, for each line, when $|\varepsilon_{d}|$ is greater than a threshold value $|\varepsilon_{d}^{c}|$, the self-consistent equations (\ref{eq14}) and (\ref{eq15}) do not have a solution. In other words, $\varepsilon_{d}^{c}$ identifies the boundary between two different phases: when $|\varepsilon_{d}|<|\varepsilon_{d}^{c}|$, the magnetic impurity is coupled to the single-layer FeSe superconductor (strong coupling regime); when $|\varepsilon_{d}|>|\varepsilon_{d}^{c}|$, the magnetic impurity and the host material are decoupled (isolated magnetic moment regime). See, e.g., Refs. [\onlinecite{Cassanello1996PRB,Cassanello1997PRB,ZhangGM2001PRL}] and [\onlinecite{Withoff1990PRL,GonzalezBuxton1998PRB,ZhuangHB2009EPL,UchoaB2012PRL}] for similar results in $d$-wave superconductors and marginal Fermi liquids. Fig. \ref{fig2}(b)-(d) show the phase diagram for three different spin-orbit couplings, $v_{\rm{so}}=0$, $12$meV{\AA} and $80$meV{\AA}. We find that the phase boundary (black lines) for the two cases with finite superconducting gaps, $v_{\rm{so}}=0$ and $v_{\rm{so}}=12$meV{\AA}, are very close to each other. The other case, $d$-wave pairing with nodes for $v_{\rm{so}}=80$meV{\AA}, has a larger strong coupling regime. In the following studies, we set the parameters located in regime (II) for all the three cases, such that the resonance states appears. 

Now we analysis the LDOS, $N_c(E,\bm{R})$. The first column in Fig. \ref{fig3} shows the LDOS close to the impurity, $N_c(E,\bm{R}=0)$, for the three typical different spin-orbit couplings we considered. The second and third columns show the corresponding spatial distributions of the resonance states in these cases at $E=\Omega_r$ and $E=-\Omega_r$, respectively (i.e., $\Omega_r/\Delta_0\approx-0.135$ in Figs. \ref{fig3}(a) and \ref{fig3}(d), $\Omega_r/\Delta_0\approx-0.115$ in Fig. \ref{fig3}(g)). By comparing both the positions and the intensities of the resonance peaks in the first column of Fig. \ref{fig3}, we find that for the two cases with finite superconducting gaps, the resonance peaks are very close to each other (the red lines), though the LDOSs for the host materials (the gray lines) are different. 
Furthermore, the spatial distributions of the first resonance states at $E=\Omega_r$ for these two cases shown in Figs. \ref{fig3}(b) and \ref{fig3}(e) are similar to each other. They both look rotation symmetric and consistent with the experimental results given in Ref. [\onlinecite{FanQ2015NatPhys}]. More detailed analysis show that the radii of the localized resonance states in Figs. \ref{fig3}(b) and \ref{fig3}(e) are about 10{\AA}, which is the same magnitude of the resonance states around a Cr adatom in single-layer FeSe superconductor as shown in [\onlinecite{FanQ2015NatPhys}]. In addition, the spatial distributions of the second resonance states for $v_{\rm{so}}=0$ and $12$meV{\AA} shown in Figs. \ref{fig3}(c) and \ref{fig3}(f) are also similar to each other. 
This result demonstrates that the LDOS is insensitive to the strength of the spin-orbit coupling as long as the superconducting gap is not closed. For the third case, $v_{\rm{so}}=80$meV{\AA}, the superconducting gap has eight nodes located in the regime between the two Fermi surfaces shown in Fig. \ref{fig1}(c). The LDOS of the quasiparticles for this case near the Fermi surface reveals linear behavior, $N_{c}(E)\propto|E|$ (See the gray line in Fig. \ref{fig3}(g) for more details). As shown in Figs. \ref{fig3}(h) and \ref{fig3}(i), the spatial distributions of LDOS corresponding to the resonance states near $E=\Omega_r$ and $E=-\Omega_r$ are fourfold rotation symmetric, which reveals the typical characteristic of $d$-wave pairing potentials.

\section{Conclusion}\label{sec4}
We have investigated the LDOS of the resonance states near a quantum magnetic impurity in single-layer FeSe superconductor using the Anderson impurity model coupled to the BdG Hamiltonian proposed in Ref. [\onlinecite{AkbariA2010PRB}]. In the strong coupling regime, the LDOS spectrum have two resonance peaks symmetrically located at the two sides of the Fermi energy. However, the intensities of the resonance peaks break the particle-hole symmetry. Three typical strengths of spin-orbit coupling are considered, i.e., (1) the vanishing spin-orbit coupling, (2) a finite spin-orbit coupling keeping the superconducting gap nodeless ($v_{\rm{so}}=12$meV{\AA}), and (3) strong spin-orbit coupling leading to the $d$-wave pairing with nodes ($v_{\rm{so}}=80$meV{\AA}). The spatial distributions of the LDOS at the resonance energy $E=\Omega_r$ for the finite-gapped cases are spatial rotation symmetric, which  behave like the tranditional plain $s$-wave pairing symmetry and consistent with the experimental results. Especially, For the second case with $v_{\rm{so}}=12$meV{\AA}, which has been named resilient nodeless $d$-wave pairing and has been used to explain the superconductivities in single-layer FeSe, our investigations give positive evidences. The third case with gap nodes displays typical behavior of the transitional $d$-wave pairing. The spatial distributions of the LDOS appear fourfold rotation symmetry.

\section*{Acknowledgment}\label{sec5}
We appreciate the support from the NSFC under Grants No. 11504106 and No. 11447167 and the Fundamental Research Funds for the Central Universities under Grant No. 2018MS049. 



\bibliographystyle{apsrev4-1}
\bibliography{G:/5_文章/jabref/magImp}

\begin{thebibliography}{54}%
\makeatletter
\providecommand \@ifxundefined [1]{%
 \@ifx{#1\undefined}
}%
\providecommand \@ifnum [1]{%
 \ifnum #1\expandafter \@firstoftwo
 \else \expandafter \@secondoftwo
 \fi
}%
\providecommand \@ifx [1]{%
 \ifx #1\expandafter \@firstoftwo
 \else \expandafter \@secondoftwo
 \fi
}%
\providecommand \natexlab [1]{#1}%
\providecommand \enquote  [1]{``#1''}%
\providecommand \bibnamefont  [1]{#1}%
\providecommand \bibfnamefont [1]{#1}%
\providecommand \citenamefont [1]{#1}%
\providecommand \href@noop [0]{\@secondoftwo}%
\providecommand \href [0]{\begingroup \@sanitize@url \@href}%
\providecommand \@href[1]{\@@startlink{#1}\@@href}%
\providecommand \@@href[1]{\endgroup#1\@@endlink}%
\providecommand \@sanitize@url [0]{\catcode `\\12\catcode `\$12\catcode
  `\&12\catcode `\#12\catcode `\^12\catcode `\_12\catcode `\%12\relax}%
\providecommand \@@startlink[1]{}%
\providecommand \@@endlink[0]{}%
\providecommand \url  [0]{\begingroup\@sanitize@url \@url }%
\providecommand \@url [1]{\endgroup\@href {#1}{\urlprefix }}%
\providecommand \urlprefix  [0]{URL }%
\providecommand \Eprint [0]{\href }%
\providecommand \doibase [0]{http://dx.doi.org/}%
\providecommand \selectlanguage [0]{\@gobble}%
\providecommand \bibinfo  [0]{\@secondoftwo}%
\providecommand \bibfield  [0]{\@secondoftwo}%
\providecommand \translation [1]{[#1]}%
\providecommand \BibitemOpen [0]{}%
\providecommand \bibitemStop [0]{}%
\providecommand \bibitemNoStop [0]{.\EOS\space}%
\providecommand \EOS [0]{\spacefactor3000\relax}%
\providecommand \BibitemShut  [1]{\csname bibitem#1\endcsname}%
\let\auto@bib@innerbib\@empty
\bibitem [{\citenamefont {Hewson}(1997)}]{Hewsonbook}%
  \BibitemOpen
  \bibfield  {author} {\bibinfo {author} {\bibfnamefont {A.~C.}\ \bibnamefont
  {Hewson}},\ }\href@noop {} {\emph {\bibinfo {title} {The Kondo Problem to
  Heavy Fermions}}}\ (\bibinfo  {publisher} {Cambridge University Press},\
  \bibinfo {year} {1997})\BibitemShut {NoStop}%
\bibitem [{\citenamefont {Balatsky}\ \emph {et~al.}(2006)\citenamefont
  {Balatsky}, \citenamefont {Vekhter},\ and\ \citenamefont
  {Zhu}}]{BalatskyAV2006RMP}%
  \BibitemOpen
  \bibfield  {author} {\bibinfo {author} {\bibfnamefont {A.~V.}\ \bibnamefont
  {Balatsky}}, \bibinfo {author} {\bibfnamefont {I.}~\bibnamefont {Vekhter}}, \
  and\ \bibinfo {author} {\bibfnamefont {J.-X.}\ \bibnamefont {Zhu}},\ }\href
  {\doibase 10.1103/RevModPhys.78.373} {\bibfield  {journal} {\bibinfo
  {journal} {Rev. Mod. Phys.}\ }\textbf {\bibinfo {volume} {78}},\ \bibinfo
  {pages} {373} (\bibinfo {year} {2006})}\BibitemShut {NoStop}%
\bibitem [{\citenamefont {Yu}(1965)}]{Yu1965APS}%
  \BibitemOpen
  \bibfield  {author} {\bibinfo {author} {\bibfnamefont {L.}~\bibnamefont
  {Yu}},\ }\href {\doibase 10.7498/aps.21.75} {\bibfield  {journal} {\bibinfo
  {journal} {Acta Physica Sinica}\ }\textbf {\bibinfo {volume} {21}},\ \bibinfo
  {eid} {75} (\bibinfo {year} {1965})}\BibitemShut {NoStop}%
\bibitem [{\citenamefont {Shiba}(1968)}]{Shiba1968CSS}%
  \BibitemOpen
  \bibfield  {author} {\bibinfo {author} {\bibfnamefont {H.}~\bibnamefont
  {Shiba}},\ }\href {\doibase 10.1143/PTP.40.435} {\bibfield  {journal}
  {\bibinfo  {journal} {Progress of Theoretical Physics}\ }\textbf {\bibinfo
  {volume} {40}},\ \bibinfo {pages} {435} (\bibinfo {year} {1968})}\BibitemShut
  {NoStop}%
\bibitem [{\citenamefont {{Rusinov}}(1969)}]{Rusinov1969JETPL}%
  \BibitemOpen
  \bibfield  {author} {\bibinfo {author} {\bibfnamefont {A.~I.}\ \bibnamefont
  {{Rusinov}}},\ }\href@noop {} {\bibfield  {journal} {\bibinfo  {journal}
  {Soviet Journal of Experimental and Theoretical Physics Letters}\ }\textbf
  {\bibinfo {volume} {9}},\ \bibinfo {pages} {85} (\bibinfo {year}
  {1969})}\BibitemShut {NoStop}%
\bibitem [{\citenamefont {Tsuei}\ \emph {et~al.}(1994)\citenamefont {Tsuei},
  \citenamefont {Kirtley}, \citenamefont {Chi}, \citenamefont {Yu-Jahnes},
  \citenamefont {Gupta}, \citenamefont {Shaw}, \citenamefont {Sun},\ and\
  \citenamefont {Ketchen}}]{Tsuei1994PRL}%
  \BibitemOpen
  \bibfield  {author} {\bibinfo {author} {\bibfnamefont {C.~C.}\ \bibnamefont
  {Tsuei}}, \bibinfo {author} {\bibfnamefont {J.~R.}\ \bibnamefont {Kirtley}},
  \bibinfo {author} {\bibfnamefont {C.~C.}\ \bibnamefont {Chi}}, \bibinfo
  {author} {\bibfnamefont {L.~S.}\ \bibnamefont {Yu-Jahnes}}, \bibinfo {author}
  {\bibfnamefont {A.}~\bibnamefont {Gupta}}, \bibinfo {author} {\bibfnamefont
  {T.}~\bibnamefont {Shaw}}, \bibinfo {author} {\bibfnamefont {J.~Z.}\
  \bibnamefont {Sun}}, \ and\ \bibinfo {author} {\bibfnamefont {M.~B.}\
  \bibnamefont {Ketchen}},\ }\href {\doibase 10.1103/PhysRevLett.73.593}
  {\bibfield  {journal} {\bibinfo  {journal} {Phys. Rev. Lett.}\ }\textbf
  {\bibinfo {volume} {73}},\ \bibinfo {pages} {593} (\bibinfo {year}
  {1994})}\BibitemShut {NoStop}%
\bibitem [{\citenamefont {Tsuei}\ \emph {et~al.}(1997)\citenamefont {Tsuei},
  \citenamefont {Kirtley}, \citenamefont {Ren}, \citenamefont {Wang},
  \citenamefont {Raffy},\ and\ \citenamefont {Li}}]{Tsuei1997Nature}%
  \BibitemOpen
  \bibfield  {author} {\bibinfo {author} {\bibfnamefont {C.~C.}\ \bibnamefont
  {Tsuei}}, \bibinfo {author} {\bibfnamefont {J.~R.}\ \bibnamefont {Kirtley}},
  \bibinfo {author} {\bibfnamefont {Z.~F.}\ \bibnamefont {Ren}}, \bibinfo
  {author} {\bibfnamefont {J.~H.}\ \bibnamefont {Wang}}, \bibinfo {author}
  {\bibfnamefont {H.}~\bibnamefont {Raffy}}, \ and\ \bibinfo {author}
  {\bibfnamefont {Z.~Z.}\ \bibnamefont {Li}},\ }\href
  {http://dx.doi.org/10.1038/387481a0} {\bibfield  {journal} {\bibinfo
  {journal} {Nature}\ }\textbf {\bibinfo {volume} {387}},\ \bibinfo {pages}
  {481 EP } (\bibinfo {year} {1997})}\BibitemShut {NoStop}%
\bibitem [{\citenamefont {Hoffman}\ \emph {et~al.}(2002)\citenamefont
  {Hoffman}, \citenamefont {McElroy}, \citenamefont {Lee}, \citenamefont
  {Lang}, \citenamefont {Eisaki}, \citenamefont {Uchida},\ and\ \citenamefont
  {Davis}}]{Hoffman2002Science}%
  \BibitemOpen
  \bibfield  {author} {\bibinfo {author} {\bibfnamefont {J.~E.}\ \bibnamefont
  {Hoffman}}, \bibinfo {author} {\bibfnamefont {K.}~\bibnamefont {McElroy}},
  \bibinfo {author} {\bibfnamefont {D.-H.}\ \bibnamefont {Lee}}, \bibinfo
  {author} {\bibfnamefont {K.~M.}\ \bibnamefont {Lang}}, \bibinfo {author}
  {\bibfnamefont {H.}~\bibnamefont {Eisaki}}, \bibinfo {author} {\bibfnamefont
  {S.}~\bibnamefont {Uchida}}, \ and\ \bibinfo {author} {\bibfnamefont {J.~C.}\
  \bibnamefont {Davis}},\ }\href {\doibase 10.1126/science.1072640} {\bibfield
  {journal} {\bibinfo  {journal} {Science}\ }\textbf {\bibinfo {volume}
  {297}},\ \bibinfo {pages} {1148} (\bibinfo {year} {2002})}\BibitemShut
  {NoStop}%
\bibitem [{\citenamefont {Wang}\ and\ \citenamefont
  {Lee}(2003)}]{WangQH2003PRB}%
  \BibitemOpen
  \bibfield  {author} {\bibinfo {author} {\bibfnamefont {Q.-H.}\ \bibnamefont
  {Wang}}\ and\ \bibinfo {author} {\bibfnamefont {D.-H.}\ \bibnamefont {Lee}},\
  }\href {\doibase 10.1103/PhysRevB.67.020511} {\bibfield  {journal} {\bibinfo
  {journal} {Phys. Rev. B}\ }\textbf {\bibinfo {volume} {67}},\ \bibinfo
  {pages} {020511} (\bibinfo {year} {2003})}\BibitemShut {NoStop}%
\bibitem [{\citenamefont {Hanaguri}\ \emph {et~al.}(2007)\citenamefont
  {Hanaguri}, \citenamefont {Kohsaka}, \citenamefont {Davis}, \citenamefont
  {Lupien}, \citenamefont {Yamada}, \citenamefont {Azuma}, \citenamefont
  {Takano}, \citenamefont {Ohishi}, \citenamefont {Ono},\ and\ \citenamefont
  {Takagi}}]{Hanaguri2007NatPhys}%
  \BibitemOpen
  \bibfield  {author} {\bibinfo {author} {\bibfnamefont {T.}~\bibnamefont
  {Hanaguri}}, \bibinfo {author} {\bibfnamefont {Y.}~\bibnamefont {Kohsaka}},
  \bibinfo {author} {\bibfnamefont {J.~C.}\ \bibnamefont {Davis}}, \bibinfo
  {author} {\bibfnamefont {C.}~\bibnamefont {Lupien}}, \bibinfo {author}
  {\bibfnamefont {I.}~\bibnamefont {Yamada}}, \bibinfo {author} {\bibfnamefont
  {M.}~\bibnamefont {Azuma}}, \bibinfo {author} {\bibfnamefont
  {M.}~\bibnamefont {Takano}}, \bibinfo {author} {\bibfnamefont
  {K.}~\bibnamefont {Ohishi}}, \bibinfo {author} {\bibfnamefont
  {M.}~\bibnamefont {Ono}}, \ and\ \bibinfo {author} {\bibfnamefont
  {H.}~\bibnamefont {Takagi}},\ }\href {http://dx.doi.org/10.1038/nphys753}
  {\bibfield  {journal} {\bibinfo  {journal} {Nature Physics}\ }\textbf
  {\bibinfo {volume} {3}},\ \bibinfo {pages} {865} (\bibinfo {year}
  {2007})}\BibitemShut {NoStop}%
\bibitem [{\citenamefont {Hanaguri}\ \emph {et~al.}(2009)\citenamefont
  {Hanaguri}, \citenamefont {Kohsaka}, \citenamefont {Ono}, \citenamefont
  {Maltseva}, \citenamefont {Coleman}, \citenamefont {Yamada}, \citenamefont
  {Azuma}, \citenamefont {Takano}, \citenamefont {Ohishi},\ and\ \citenamefont
  {Takagi}}]{Hanaguri2009Science}%
  \BibitemOpen
  \bibfield  {author} {\bibinfo {author} {\bibfnamefont {T.}~\bibnamefont
  {Hanaguri}}, \bibinfo {author} {\bibfnamefont {Y.}~\bibnamefont {Kohsaka}},
  \bibinfo {author} {\bibfnamefont {M.}~\bibnamefont {Ono}}, \bibinfo {author}
  {\bibfnamefont {M.}~\bibnamefont {Maltseva}}, \bibinfo {author}
  {\bibfnamefont {P.}~\bibnamefont {Coleman}}, \bibinfo {author} {\bibfnamefont
  {I.}~\bibnamefont {Yamada}}, \bibinfo {author} {\bibfnamefont
  {M.}~\bibnamefont {Azuma}}, \bibinfo {author} {\bibfnamefont
  {M.}~\bibnamefont {Takano}}, \bibinfo {author} {\bibfnamefont
  {K.}~\bibnamefont {Ohishi}}, \ and\ \bibinfo {author} {\bibfnamefont
  {H.}~\bibnamefont {Takagi}},\ }\href {\doibase 10.1126/science.1166138}
  {\bibfield  {journal} {\bibinfo  {journal} {Science}\ }\textbf {\bibinfo
  {volume} {323}},\ \bibinfo {pages} {923} (\bibinfo {year}
  {2009})}\BibitemShut {NoStop}%
\bibitem [{\citenamefont {Pan}\ \emph {et~al.}(2000)\citenamefont {Pan},
  \citenamefont {Hudson}, \citenamefont {Lang}, \citenamefont {Eisaki},
  \citenamefont {Uchida},\ and\ \citenamefont {Davis}}]{PanSH2000Nat}%
  \BibitemOpen
  \bibfield  {author} {\bibinfo {author} {\bibfnamefont {S.~H.}\ \bibnamefont
  {Pan}}, \bibinfo {author} {\bibfnamefont {E.~W.}\ \bibnamefont {Hudson}},
  \bibinfo {author} {\bibfnamefont {K.~M.}\ \bibnamefont {Lang}}, \bibinfo
  {author} {\bibfnamefont {H.}~\bibnamefont {Eisaki}}, \bibinfo {author}
  {\bibfnamefont {S.}~\bibnamefont {Uchida}}, \ and\ \bibinfo {author}
  {\bibfnamefont {J.~C.}\ \bibnamefont {Davis}},\ }\href
  {http://dx.doi.org/10.1038/35001534} {\bibfield  {journal} {\bibinfo
  {journal} {Nature}\ }\textbf {\bibinfo {volume} {403}},\ \bibinfo {pages}
  {746} (\bibinfo {year} {2000})}\BibitemShut {NoStop}%
\bibitem [{\citenamefont {Hudson}\ \emph {et~al.}(2001)\citenamefont {Hudson},
  \citenamefont {Lang}, \citenamefont {Madhavan}, \citenamefont {Pan},
  \citenamefont {Eisaki}, \citenamefont {Uchida},\ and\ \citenamefont
  {Davis}}]{HudsonEW2001Nat}%
  \BibitemOpen
  \bibfield  {author} {\bibinfo {author} {\bibfnamefont {E.~W.}\ \bibnamefont
  {Hudson}}, \bibinfo {author} {\bibfnamefont {K.~M.}\ \bibnamefont {Lang}},
  \bibinfo {author} {\bibfnamefont {V.}~\bibnamefont {Madhavan}}, \bibinfo
  {author} {\bibfnamefont {S.~H.}\ \bibnamefont {Pan}}, \bibinfo {author}
  {\bibfnamefont {H.}~\bibnamefont {Eisaki}}, \bibinfo {author} {\bibfnamefont
  {S.}~\bibnamefont {Uchida}}, \ and\ \bibinfo {author} {\bibfnamefont {J.~C.}\
  \bibnamefont {Davis}},\ }\href {http://dx.doi.org/10.1038/35082019}
  {\bibfield  {journal} {\bibinfo  {journal} {Nature}\ }\textbf {\bibinfo
  {volume} {411}},\ \bibinfo {pages} {920} (\bibinfo {year}
  {2001})}\BibitemShut {NoStop}%
\bibitem [{\citenamefont {Bobroff}\ \emph {et~al.}(2001)\citenamefont
  {Bobroff}, \citenamefont {Alloul}, \citenamefont {MacFarlane}, \citenamefont
  {Mendels}, \citenamefont {Blanchard}, \citenamefont {Collin},\ and\
  \citenamefont {Marucco}}]{BobroffJ2001PRL}%
  \BibitemOpen
  \bibfield  {author} {\bibinfo {author} {\bibfnamefont {J.}~\bibnamefont
  {Bobroff}}, \bibinfo {author} {\bibfnamefont {H.}~\bibnamefont {Alloul}},
  \bibinfo {author} {\bibfnamefont {W.~A.}\ \bibnamefont {MacFarlane}},
  \bibinfo {author} {\bibfnamefont {P.}~\bibnamefont {Mendels}}, \bibinfo
  {author} {\bibfnamefont {N.}~\bibnamefont {Blanchard}}, \bibinfo {author}
  {\bibfnamefont {G.}~\bibnamefont {Collin}}, \ and\ \bibinfo {author}
  {\bibfnamefont {J.-F.}\ \bibnamefont {Marucco}},\ }\href {\doibase
  10.1103/PhysRevLett.86.4116} {\bibfield  {journal} {\bibinfo  {journal}
  {Phys. Rev. Lett.}\ }\textbf {\bibinfo {volume} {86}},\ \bibinfo {pages}
  {4116} (\bibinfo {year} {2001})}\BibitemShut {NoStop}%
\bibitem [{\citenamefont {Zhang}\ \emph {et~al.}(2001)\citenamefont {Zhang},
  \citenamefont {Hu},\ and\ \citenamefont {Yu}}]{ZhangGM2001PRL}%
  \BibitemOpen
  \bibfield  {author} {\bibinfo {author} {\bibfnamefont {G.-M.}\ \bibnamefont
  {Zhang}}, \bibinfo {author} {\bibfnamefont {H.}~\bibnamefont {Hu}}, \ and\
  \bibinfo {author} {\bibfnamefont {L.}~\bibnamefont {Yu}},\ }\href {\doibase
  10.1103/PhysRevLett.86.704} {\bibfield  {journal} {\bibinfo  {journal} {Phys.
  Rev. Lett.}\ }\textbf {\bibinfo {volume} {86}},\ \bibinfo {pages} {704}
  (\bibinfo {year} {2001})}\BibitemShut {NoStop}%
\bibitem [{\citenamefont {Polkovnikov}\ \emph {et~al.}(2001)\citenamefont
  {Polkovnikov}, \citenamefont {Sachdev},\ and\ \citenamefont
  {Vojta}}]{Polkovnikov2001PRL}%
  \BibitemOpen
  \bibfield  {author} {\bibinfo {author} {\bibfnamefont {A.}~\bibnamefont
  {Polkovnikov}}, \bibinfo {author} {\bibfnamefont {S.}~\bibnamefont
  {Sachdev}}, \ and\ \bibinfo {author} {\bibfnamefont {M.}~\bibnamefont
  {Vojta}},\ }\href {\doibase 10.1103/PhysRevLett.86.296} {\bibfield  {journal}
  {\bibinfo  {journal} {Phys. Rev. Lett.}\ }\textbf {\bibinfo {volume} {86}},\
  \bibinfo {pages} {296} (\bibinfo {year} {2001})}\BibitemShut {NoStop}%
\bibitem [{\citenamefont {Zhu}\ and\ \citenamefont
  {Ting}(2000)}]{ZhuJX2000PRB}%
  \BibitemOpen
  \bibfield  {author} {\bibinfo {author} {\bibfnamefont {J.-X.}\ \bibnamefont
  {Zhu}}\ and\ \bibinfo {author} {\bibfnamefont {C.~S.}\ \bibnamefont {Ting}},\
  }\href {\doibase 10.1103/PhysRevB.63.020506} {\bibfield  {journal} {\bibinfo
  {journal} {Phys. Rev. B}\ }\textbf {\bibinfo {volume} {63}},\ \bibinfo
  {pages} {020506} (\bibinfo {year} {2000})}\BibitemShut {NoStop}%
\bibitem [{\citenamefont {Vojta}\ \emph {et~al.}(2002)\citenamefont {Vojta},
  \citenamefont {Zitzler}, \citenamefont {Bulla},\ and\ \citenamefont
  {Pruschke}}]{VojtaM2002PRB}%
  \BibitemOpen
  \bibfield  {author} {\bibinfo {author} {\bibfnamefont {M.}~\bibnamefont
  {Vojta}}, \bibinfo {author} {\bibfnamefont {R.}~\bibnamefont {Zitzler}},
  \bibinfo {author} {\bibfnamefont {R.}~\bibnamefont {Bulla}}, \ and\ \bibinfo
  {author} {\bibfnamefont {T.}~\bibnamefont {Pruschke}},\ }\href {\doibase
  10.1103/PhysRevB.66.134527} {\bibfield  {journal} {\bibinfo  {journal} {Phys.
  Rev. B}\ }\textbf {\bibinfo {volume} {66}},\ \bibinfo {pages} {134527}
  (\bibinfo {year} {2002})}\BibitemShut {NoStop}%
\bibitem [{\citenamefont {Polkovnikov}(2002)}]{Polkovnikov2002PRB}%
  \BibitemOpen
  \bibfield  {author} {\bibinfo {author} {\bibfnamefont {A.}~\bibnamefont
  {Polkovnikov}},\ }\href {\doibase 10.1103/PhysRevB.65.064503} {\bibfield
  {journal} {\bibinfo  {journal} {Phys. Rev. B}\ }\textbf {\bibinfo {volume}
  {65}},\ \bibinfo {pages} {064503} (\bibinfo {year} {2002})}\BibitemShut
  {NoStop}%
\bibitem [{\citenamefont {Dai}\ and\ \citenamefont {Wang}(2003)}]{DaiX2003PRB}%
  \BibitemOpen
  \bibfield  {author} {\bibinfo {author} {\bibfnamefont {X.}~\bibnamefont
  {Dai}}\ and\ \bibinfo {author} {\bibfnamefont {Z.}~\bibnamefont {Wang}},\
  }\href {\doibase 10.1103/PhysRevB.67.180507} {\bibfield  {journal} {\bibinfo
  {journal} {Phys. Rev. B}\ }\textbf {\bibinfo {volume} {67}},\ \bibinfo
  {pages} {180507} (\bibinfo {year} {2003})}\BibitemShut {NoStop}%
\bibitem [{\citenamefont {Baar}\ \emph {et~al.}(2016)\citenamefont {Baar},
  \citenamefont {Momono}, \citenamefont {Kawamura}, \citenamefont {Kobayashi},
  \citenamefont {Iwasaki}, \citenamefont {Sakawaki}, \citenamefont {Amakai},
  \citenamefont {Takano}, \citenamefont {Kurosawa}, \citenamefont {Oda},\ and\
  \citenamefont {Ido}}]{BaarS2016JSNM}%
  \BibitemOpen
  \bibfield  {author} {\bibinfo {author} {\bibfnamefont {S.}~\bibnamefont
  {Baar}}, \bibinfo {author} {\bibfnamefont {N.}~\bibnamefont {Momono}},
  \bibinfo {author} {\bibfnamefont {K.}~\bibnamefont {Kawamura}}, \bibinfo
  {author} {\bibfnamefont {Y.}~\bibnamefont {Kobayashi}}, \bibinfo {author}
  {\bibfnamefont {S.}~\bibnamefont {Iwasaki}}, \bibinfo {author} {\bibfnamefont
  {T.}~\bibnamefont {Sakawaki}}, \bibinfo {author} {\bibfnamefont
  {Y.}~\bibnamefont {Amakai}}, \bibinfo {author} {\bibfnamefont
  {H.}~\bibnamefont {Takano}}, \bibinfo {author} {\bibfnamefont
  {T.}~\bibnamefont {Kurosawa}}, \bibinfo {author} {\bibfnamefont
  {M.}~\bibnamefont {Oda}}, \ and\ \bibinfo {author} {\bibfnamefont
  {M.}~\bibnamefont {Ido}},\ }\href {\doibase 10.1007/s10948-015-3320-2}
  {\bibfield  {journal} {\bibinfo  {journal} {Journal of Superconductivity and
  Novel Magnetism}\ }\textbf {\bibinfo {volume} {29}},\ \bibinfo {pages} {659}
  (\bibinfo {year} {2016})}\BibitemShut {NoStop}%
\bibitem [{\citenamefont {Tsai}\ \emph {et~al.}(2009)\citenamefont {Tsai},
  \citenamefont {Zhang}, \citenamefont {Fang},\ and\ \citenamefont
  {Hu}}]{TsaiWF2009PRB}%
  \BibitemOpen
  \bibfield  {author} {\bibinfo {author} {\bibfnamefont {W.-F.}\ \bibnamefont
  {Tsai}}, \bibinfo {author} {\bibfnamefont {Y.-Y.}\ \bibnamefont {Zhang}},
  \bibinfo {author} {\bibfnamefont {C.}~\bibnamefont {Fang}}, \ and\ \bibinfo
  {author} {\bibfnamefont {J.}~\bibnamefont {Hu}},\ }\href {\doibase
  10.1103/PhysRevB.80.064513} {\bibfield  {journal} {\bibinfo  {journal} {Phys.
  Rev. B}\ }\textbf {\bibinfo {volume} {80}},\ \bibinfo {pages} {064513}
  (\bibinfo {year} {2009})}\BibitemShut {NoStop}%
\bibitem [{\citenamefont {Bang}\ \emph {et~al.}(2009)\citenamefont {Bang},
  \citenamefont {Choi},\ and\ \citenamefont {Won}}]{BangY2009PRB}%
  \BibitemOpen
  \bibfield  {author} {\bibinfo {author} {\bibfnamefont {Y.}~\bibnamefont
  {Bang}}, \bibinfo {author} {\bibfnamefont {H.-Y.}\ \bibnamefont {Choi}}, \
  and\ \bibinfo {author} {\bibfnamefont {H.}~\bibnamefont {Won}},\ }\href
  {\doibase 10.1103/PhysRevB.79.054529} {\bibfield  {journal} {\bibinfo
  {journal} {Phys. Rev. B}\ }\textbf {\bibinfo {volume} {79}},\ \bibinfo
  {pages} {054529} (\bibinfo {year} {2009})}\BibitemShut {NoStop}%
\bibitem [{\citenamefont {Akbari}\ \emph {et~al.}(2010)\citenamefont {Akbari},
  \citenamefont {Eremin},\ and\ \citenamefont {Thalmeier}}]{AkbariA2010PRB}%
  \BibitemOpen
  \bibfield  {author} {\bibinfo {author} {\bibfnamefont {A.}~\bibnamefont
  {Akbari}}, \bibinfo {author} {\bibfnamefont {I.}~\bibnamefont {Eremin}}, \
  and\ \bibinfo {author} {\bibfnamefont {P.}~\bibnamefont {Thalmeier}},\ }\href
  {\doibase 10.1103/PhysRevB.81.014524} {\bibfield  {journal} {\bibinfo
  {journal} {Phys. Rev. B}\ }\textbf {\bibinfo {volume} {81}},\ \bibinfo
  {pages} {014524} (\bibinfo {year} {2010})}\BibitemShut {NoStop}%
\bibitem [{\citenamefont {Sau}\ and\ \citenamefont
  {Demler}(2013)}]{SauJD2013PRB}%
  \BibitemOpen
  \bibfield  {author} {\bibinfo {author} {\bibfnamefont {J.~D.}\ \bibnamefont
  {Sau}}\ and\ \bibinfo {author} {\bibfnamefont {E.}~\bibnamefont {Demler}},\
  }\href {\doibase 10.1103/PhysRevB.88.205402} {\bibfield  {journal} {\bibinfo
  {journal} {Phys. Rev. B}\ }\textbf {\bibinfo {volume} {88}},\ \bibinfo
  {pages} {205402} (\bibinfo {year} {2013})}\BibitemShut {NoStop}%
\bibitem [{\citenamefont {Fu}\ \emph {et~al.}(2012)\citenamefont {Fu},
  \citenamefont {Zhang}, \citenamefont {Wang},\ and\ \citenamefont
  {Li}}]{FuZG2012JPCM}%
  \BibitemOpen
  \bibfield  {author} {\bibinfo {author} {\bibfnamefont {Z.-G.}\ \bibnamefont
  {Fu}}, \bibinfo {author} {\bibfnamefont {P.}~\bibnamefont {Zhang}}, \bibinfo
  {author} {\bibfnamefont {Z.}~\bibnamefont {Wang}}, \ and\ \bibinfo {author}
  {\bibfnamefont {S.-S.}\ \bibnamefont {Li}},\ }\href
  {http://stacks.iop.org/0953-8984/24/i=14/a=145502} {\bibfield  {journal}
  {\bibinfo  {journal} {Journal of Physics: Condensed Matter}\ }\textbf
  {\bibinfo {volume} {24}},\ \bibinfo {pages} {145502} (\bibinfo {year}
  {2012})}\BibitemShut {NoStop}%
\bibitem [{\citenamefont {Zha}\ and\ \citenamefont {Jin}(2017)}]{ZhaGQ2017EPL}%
  \BibitemOpen
  \bibfield  {author} {\bibinfo {author} {\bibfnamefont {G.-Q.}\ \bibnamefont
  {Zha}}\ and\ \bibinfo {author} {\bibfnamefont {Y.-Y.}\ \bibnamefont {Jin}},\
  }\href {http://stacks.iop.org/0295-5075/120/i=2/a=27002} {\bibfield
  {journal} {\bibinfo  {journal} {EPL (Europhysics Letters)}\ }\textbf
  {\bibinfo {volume} {120}},\ \bibinfo {pages} {27002} (\bibinfo {year}
  {2017})}\BibitemShut {NoStop}%
\bibitem [{\citenamefont {Guo}\ \emph {et~al.}(2017)\citenamefont {Guo},
  \citenamefont {Li},\ and\ \citenamefont {Chen}}]{GuoYW2017FP}%
  \BibitemOpen
  \bibfield  {author} {\bibinfo {author} {\bibfnamefont {Y.-W.}\ \bibnamefont
  {Guo}}, \bibinfo {author} {\bibfnamefont {W.}~\bibnamefont {Li}}, \ and\
  \bibinfo {author} {\bibfnamefont {Y.}~\bibnamefont {Chen}},\ }\href {\doibase
  10.1007/s11467-017-0683-9} {\bibfield  {journal} {\bibinfo  {journal}
  {Frontiers of Physics}\ }\textbf {\bibinfo {volume} {12}},\ \bibinfo {pages}
  {127403} (\bibinfo {year} {2017})}\BibitemShut {NoStop}%
\bibitem [{\citenamefont {Hsu}\ \emph {et~al.}(2008)\citenamefont {Hsu},
  \citenamefont {Luo}, \citenamefont {Yeh}, \citenamefont {Chen}, \citenamefont
  {Huang}, \citenamefont {Wu}, \citenamefont {Lee}, \citenamefont {Huang},
  \citenamefont {Chu}, \citenamefont {Yan},\ and\ \citenamefont
  {Wu}}]{HsuFC2008PNAS}%
  \BibitemOpen
  \bibfield  {author} {\bibinfo {author} {\bibfnamefont {F.-C.}\ \bibnamefont
  {Hsu}}, \bibinfo {author} {\bibfnamefont {J.-Y.}\ \bibnamefont {Luo}},
  \bibinfo {author} {\bibfnamefont {K.-W.}\ \bibnamefont {Yeh}}, \bibinfo
  {author} {\bibfnamefont {T.-K.}\ \bibnamefont {Chen}}, \bibinfo {author}
  {\bibfnamefont {T.-W.}\ \bibnamefont {Huang}}, \bibinfo {author}
  {\bibfnamefont {P.~M.}\ \bibnamefont {Wu}}, \bibinfo {author} {\bibfnamefont
  {Y.-C.}\ \bibnamefont {Lee}}, \bibinfo {author} {\bibfnamefont {Y.-L.}\
  \bibnamefont {Huang}}, \bibinfo {author} {\bibfnamefont {Y.-Y.}\ \bibnamefont
  {Chu}}, \bibinfo {author} {\bibfnamefont {D.-C.}\ \bibnamefont {Yan}}, \ and\
  \bibinfo {author} {\bibfnamefont {M.-K.}\ \bibnamefont {Wu}},\ }\href
  {\doibase 10.1073/pnas.0807325105} {\bibfield  {journal} {\bibinfo  {journal}
  {Proceedings of the National Academy of Sciences}\ }\textbf {\bibinfo
  {volume} {105}},\ \bibinfo {pages} {14262} (\bibinfo {year}
  {2008})}\BibitemShut {NoStop}%
\bibitem [{\citenamefont {Wang}\ \emph {et~al.}(2015)\citenamefont {Wang},
  \citenamefont {Shen}, \citenamefont {Pan}, \citenamefont {Hao}, \citenamefont
  {Ma}, \citenamefont {Zhou}, \citenamefont {Steffens}, \citenamefont
  {Schmalzl}, \citenamefont {Forrest}, \citenamefont {Abdel-Hafiez},
  \citenamefont {Chen}, \citenamefont {Chareev}, \citenamefont {Vasiliev},
  \citenamefont {Bourges}, \citenamefont {Sidis}, \citenamefont {Cao},\ and\
  \citenamefont {Zhao}}]{WangQNatMat2015}%
  \BibitemOpen
  \bibfield  {author} {\bibinfo {author} {\bibfnamefont {Q.}~\bibnamefont
  {Wang}}, \bibinfo {author} {\bibfnamefont {Y.}~\bibnamefont {Shen}}, \bibinfo
  {author} {\bibfnamefont {B.}~\bibnamefont {Pan}}, \bibinfo {author}
  {\bibfnamefont {Y.}~\bibnamefont {Hao}}, \bibinfo {author} {\bibfnamefont
  {M.}~\bibnamefont {Ma}}, \bibinfo {author} {\bibfnamefont {F.}~\bibnamefont
  {Zhou}}, \bibinfo {author} {\bibfnamefont {P.}~\bibnamefont {Steffens}},
  \bibinfo {author} {\bibfnamefont {K.}~\bibnamefont {Schmalzl}}, \bibinfo
  {author} {\bibfnamefont {T.~R.}\ \bibnamefont {Forrest}}, \bibinfo {author}
  {\bibfnamefont {M.}~\bibnamefont {Abdel-Hafiez}}, \bibinfo {author}
  {\bibfnamefont {X.}~\bibnamefont {Chen}}, \bibinfo {author} {\bibfnamefont
  {D.~A.}\ \bibnamefont {Chareev}}, \bibinfo {author} {\bibfnamefont {A.~N.}\
  \bibnamefont {Vasiliev}}, \bibinfo {author} {\bibfnamefont {P.}~\bibnamefont
  {Bourges}}, \bibinfo {author} {\bibfnamefont {Y.}~\bibnamefont {Sidis}},
  \bibinfo {author} {\bibfnamefont {H.}~\bibnamefont {Cao}}, \ and\ \bibinfo
  {author} {\bibfnamefont {J.}~\bibnamefont {Zhao}},\ }\href
  {http://dx.doi.org/10.1038/nmat4492} {\bibfield  {journal} {\bibinfo
  {journal} {Nature Materials}\ }\textbf {\bibinfo {volume} {15}} (\bibinfo
  {year} {2015})}\BibitemShut {NoStop}%
\bibitem [{\citenamefont {Wang}\ \emph {et~al.}(2012)\citenamefont {Wang},
  \citenamefont {Li}, \citenamefont {Zhang}, \citenamefont {Zhang},
  \citenamefont {Zhang}, \citenamefont {Li}, \citenamefont {Ding},
  \citenamefont {Ou}, \citenamefont {Deng}, \citenamefont {Chang},
  \citenamefont {Wen}, \citenamefont {Song}, \citenamefont {He}, \citenamefont
  {Jia}, \citenamefont {Ji}, \citenamefont {Wang}, \citenamefont {Wang},
  \citenamefont {Chen}, \citenamefont {Ma},\ and\ \citenamefont
  {Xue}}]{WangQY2012CPL}%
  \BibitemOpen
  \bibfield  {author} {\bibinfo {author} {\bibfnamefont {Q.-Y.}\ \bibnamefont
  {Wang}}, \bibinfo {author} {\bibfnamefont {Z.}~\bibnamefont {Li}}, \bibinfo
  {author} {\bibfnamefont {W.-H.}\ \bibnamefont {Zhang}}, \bibinfo {author}
  {\bibfnamefont {Z.-C.}\ \bibnamefont {Zhang}}, \bibinfo {author}
  {\bibfnamefont {J.-S.}\ \bibnamefont {Zhang}}, \bibinfo {author}
  {\bibfnamefont {W.}~\bibnamefont {Li}}, \bibinfo {author} {\bibfnamefont
  {H.}~\bibnamefont {Ding}}, \bibinfo {author} {\bibfnamefont {Y.-B.}\
  \bibnamefont {Ou}}, \bibinfo {author} {\bibfnamefont {P.}~\bibnamefont
  {Deng}}, \bibinfo {author} {\bibfnamefont {K.}~\bibnamefont {Chang}},
  \bibinfo {author} {\bibfnamefont {J.}~\bibnamefont {Wen}}, \bibinfo {author}
  {\bibfnamefont {C.-L.}\ \bibnamefont {Song}}, \bibinfo {author}
  {\bibfnamefont {K.}~\bibnamefont {He}}, \bibinfo {author} {\bibfnamefont
  {J.-F.}\ \bibnamefont {Jia}}, \bibinfo {author} {\bibfnamefont {S.-H.}\
  \bibnamefont {Ji}}, \bibinfo {author} {\bibfnamefont {Y.-Y.}\ \bibnamefont
  {Wang}}, \bibinfo {author} {\bibfnamefont {L.-L.}\ \bibnamefont {Wang}},
  \bibinfo {author} {\bibfnamefont {X.}~\bibnamefont {Chen}}, \bibinfo {author}
  {\bibfnamefont {X.-C.}\ \bibnamefont {Ma}}, \ and\ \bibinfo {author}
  {\bibfnamefont {Q.-K.}\ \bibnamefont {Xue}},\ }\href
  {http://stacks.iop.org/0256-307X/29/i=3/a=037402} {\bibfield  {journal}
  {\bibinfo  {journal} {Chinese Physics Letters}\ }\textbf {\bibinfo {volume}
  {29}},\ \bibinfo {pages} {037402} (\bibinfo {year} {2012})}\BibitemShut
  {NoStop}%
\bibitem [{\citenamefont {Liu}\ \emph {et~al.}(2012)\citenamefont {Liu},
  \citenamefont {Zhang}, \citenamefont {Mou}, \citenamefont {He}, \citenamefont
  {Ou}, \citenamefont {Wang}, \citenamefont {Li}, \citenamefont {Wang},
  \citenamefont {Zhao}, \citenamefont {He}, \citenamefont {Peng}, \citenamefont
  {Liu}, \citenamefont {Chen}, \citenamefont {Yu}, \citenamefont {Liu},
  \citenamefont {Dong}, \citenamefont {Zhang}, \citenamefont {Chen},
  \citenamefont {Xu}, \citenamefont {Hu}, \citenamefont {Chen}, \citenamefont
  {Ma}, \citenamefont {Xue},\ and\ \citenamefont {Zhou}}]{Liu2012NatComm}%
  \BibitemOpen
  \bibfield  {author} {\bibinfo {author} {\bibfnamefont {D.}~\bibnamefont
  {Liu}}, \bibinfo {author} {\bibfnamefont {W.}~\bibnamefont {Zhang}}, \bibinfo
  {author} {\bibfnamefont {D.}~\bibnamefont {Mou}}, \bibinfo {author}
  {\bibfnamefont {J.}~\bibnamefont {He}}, \bibinfo {author} {\bibfnamefont
  {Y.-B.}\ \bibnamefont {Ou}}, \bibinfo {author} {\bibfnamefont {Q.-Y.}\
  \bibnamefont {Wang}}, \bibinfo {author} {\bibfnamefont {Z.}~\bibnamefont
  {Li}}, \bibinfo {author} {\bibfnamefont {L.}~\bibnamefont {Wang}}, \bibinfo
  {author} {\bibfnamefont {L.}~\bibnamefont {Zhao}}, \bibinfo {author}
  {\bibfnamefont {S.}~\bibnamefont {He}}, \bibinfo {author} {\bibfnamefont
  {Y.}~\bibnamefont {Peng}}, \bibinfo {author} {\bibfnamefont {X.}~\bibnamefont
  {Liu}}, \bibinfo {author} {\bibfnamefont {C.}~\bibnamefont {Chen}}, \bibinfo
  {author} {\bibfnamefont {L.}~\bibnamefont {Yu}}, \bibinfo {author}
  {\bibfnamefont {G.}~\bibnamefont {Liu}}, \bibinfo {author} {\bibfnamefont
  {X.}~\bibnamefont {Dong}}, \bibinfo {author} {\bibfnamefont {J.}~\bibnamefont
  {Zhang}}, \bibinfo {author} {\bibfnamefont {C.}~\bibnamefont {Chen}},
  \bibinfo {author} {\bibfnamefont {Z.}~\bibnamefont {Xu}}, \bibinfo {author}
  {\bibfnamefont {J.}~\bibnamefont {Hu}}, \bibinfo {author} {\bibfnamefont
  {X.}~\bibnamefont {Chen}}, \bibinfo {author} {\bibfnamefont {X.}~\bibnamefont
  {Ma}}, \bibinfo {author} {\bibfnamefont {Q.}~\bibnamefont {Xue}}, \ and\
  \bibinfo {author} {\bibfnamefont {X.~J.}\ \bibnamefont {Zhou}},\ }\href
  {http://dx.doi.org/10.1038/ncomms1946} {\bibfield  {journal} {\bibinfo
  {journal} {Nature Communications}\ }\textbf {\bibinfo {volume} {3}},\
  \bibinfo {pages} {931} (\bibinfo {year} {2012})}\BibitemShut {NoStop}%
\bibitem [{\citenamefont {He}\ \emph {et~al.}(2013)\citenamefont {He},
  \citenamefont {He}, \citenamefont {Zhang}, \citenamefont {Zhao},
  \citenamefont {Liu}, \citenamefont {Liu}, \citenamefont {Mou}, \citenamefont
  {Ou}, \citenamefont {Wang}, \citenamefont {Li}, \citenamefont {Wang},
  \citenamefont {Peng}, \citenamefont {Liu}, \citenamefont {Chen},
  \citenamefont {Yu}, \citenamefont {Liu}, \citenamefont {Dong}, \citenamefont
  {Zhang}, \citenamefont {Chen}, \citenamefont {Xu}, \citenamefont {Chen},
  \citenamefont {Ma}, \citenamefont {Xue},\ and\ \citenamefont
  {Zhou}}]{HeSL2013NatMat}%
  \BibitemOpen
  \bibfield  {author} {\bibinfo {author} {\bibfnamefont {S.}~\bibnamefont
  {He}}, \bibinfo {author} {\bibfnamefont {J.}~\bibnamefont {He}}, \bibinfo
  {author} {\bibfnamefont {W.}~\bibnamefont {Zhang}}, \bibinfo {author}
  {\bibfnamefont {L.}~\bibnamefont {Zhao}}, \bibinfo {author} {\bibfnamefont
  {D.}~\bibnamefont {Liu}}, \bibinfo {author} {\bibfnamefont {X.}~\bibnamefont
  {Liu}}, \bibinfo {author} {\bibfnamefont {D.}~\bibnamefont {Mou}}, \bibinfo
  {author} {\bibfnamefont {Y.-B.}\ \bibnamefont {Ou}}, \bibinfo {author}
  {\bibfnamefont {Q.-Y.}\ \bibnamefont {Wang}}, \bibinfo {author}
  {\bibfnamefont {Z.}~\bibnamefont {Li}}, \bibinfo {author} {\bibfnamefont
  {L.}~\bibnamefont {Wang}}, \bibinfo {author} {\bibfnamefont {Y.}~\bibnamefont
  {Peng}}, \bibinfo {author} {\bibfnamefont {Y.}~\bibnamefont {Liu}}, \bibinfo
  {author} {\bibfnamefont {C.}~\bibnamefont {Chen}}, \bibinfo {author}
  {\bibfnamefont {L.}~\bibnamefont {Yu}}, \bibinfo {author} {\bibfnamefont
  {G.}~\bibnamefont {Liu}}, \bibinfo {author} {\bibfnamefont {X.}~\bibnamefont
  {Dong}}, \bibinfo {author} {\bibfnamefont {J.}~\bibnamefont {Zhang}},
  \bibinfo {author} {\bibfnamefont {C.}~\bibnamefont {Chen}}, \bibinfo {author}
  {\bibfnamefont {Z.}~\bibnamefont {Xu}}, \bibinfo {author} {\bibfnamefont
  {X.}~\bibnamefont {Chen}}, \bibinfo {author} {\bibfnamefont {X.}~\bibnamefont
  {Ma}}, \bibinfo {author} {\bibfnamefont {Q.}~\bibnamefont {Xue}}, \ and\
  \bibinfo {author} {\bibfnamefont {X.~J.}\ \bibnamefont {Zhou}},\ }\href
  {http://dx.doi.org/10.1038/nmat3648} {\bibfield  {journal} {\bibinfo
  {journal} {Nature Materials}\ }\textbf {\bibinfo {volume} {12}},\ \bibinfo
  {pages} {605} (\bibinfo {year} {2013})}\BibitemShut {NoStop}%
\bibitem [{\citenamefont {Tan}\ \emph {et~al.}(2013)\citenamefont {Tan},
  \citenamefont {Zhang}, \citenamefont {Xia}, \citenamefont {Ye}, \citenamefont
  {Chen}, \citenamefont {Xie}, \citenamefont {Peng}, \citenamefont {Xu},
  \citenamefont {Fan}, \citenamefont {Xu}, \citenamefont {Jiang}, \citenamefont
  {Zhang}, \citenamefont {Lai}, \citenamefont {Xiang}, \citenamefont {Hu},
  \citenamefont {Xie},\ and\ \citenamefont {Feng}}]{TanSY2013NatMat}%
  \BibitemOpen
  \bibfield  {author} {\bibinfo {author} {\bibfnamefont {S.}~\bibnamefont
  {Tan}}, \bibinfo {author} {\bibfnamefont {Y.}~\bibnamefont {Zhang}}, \bibinfo
  {author} {\bibfnamefont {M.}~\bibnamefont {Xia}}, \bibinfo {author}
  {\bibfnamefont {Z.}~\bibnamefont {Ye}}, \bibinfo {author} {\bibfnamefont
  {F.}~\bibnamefont {Chen}}, \bibinfo {author} {\bibfnamefont {X.}~\bibnamefont
  {Xie}}, \bibinfo {author} {\bibfnamefont {R.}~\bibnamefont {Peng}}, \bibinfo
  {author} {\bibfnamefont {D.}~\bibnamefont {Xu}}, \bibinfo {author}
  {\bibfnamefont {Q.}~\bibnamefont {Fan}}, \bibinfo {author} {\bibfnamefont
  {H.}~\bibnamefont {Xu}}, \bibinfo {author} {\bibfnamefont {J.}~\bibnamefont
  {Jiang}}, \bibinfo {author} {\bibfnamefont {T.}~\bibnamefont {Zhang}},
  \bibinfo {author} {\bibfnamefont {X.}~\bibnamefont {Lai}}, \bibinfo {author}
  {\bibfnamefont {T.}~\bibnamefont {Xiang}}, \bibinfo {author} {\bibfnamefont
  {J.}~\bibnamefont {Hu}}, \bibinfo {author} {\bibfnamefont {B.}~\bibnamefont
  {Xie}}, \ and\ \bibinfo {author} {\bibfnamefont {D.}~\bibnamefont {Feng}},\
  }\href {http://dx.doi.org/10.1038/nmat3654} {\bibfield  {journal} {\bibinfo
  {journal} {Nature Materials}\ }\textbf {\bibinfo {volume} {12}},\ \bibinfo
  {pages} {634} (\bibinfo {year} {2013})}\BibitemShut {NoStop}%
\bibitem [{\citenamefont {Lee}\ \emph {et~al.}(2014)\citenamefont {Lee},
  \citenamefont {Schmitt}, \citenamefont {Moore}, \citenamefont {Johnston},
  \citenamefont {Cui}, \citenamefont {Li}, \citenamefont {Yi}, \citenamefont
  {Liu}, \citenamefont {Hashimoto}, \citenamefont {Zhang}, \citenamefont {Lu},
  \citenamefont {Devereaux}, \citenamefont {Lee},\ and\ \citenamefont
  {Shen}}]{LeeJJ2014NatLett}%
  \BibitemOpen
  \bibfield  {author} {\bibinfo {author} {\bibfnamefont {J.~J.}\ \bibnamefont
  {Lee}}, \bibinfo {author} {\bibfnamefont {F.~T.}\ \bibnamefont {Schmitt}},
  \bibinfo {author} {\bibfnamefont {R.~G.}\ \bibnamefont {Moore}}, \bibinfo
  {author} {\bibfnamefont {S.}~\bibnamefont {Johnston}}, \bibinfo {author}
  {\bibfnamefont {Y.-T.}\ \bibnamefont {Cui}}, \bibinfo {author} {\bibfnamefont
  {W.}~\bibnamefont {Li}}, \bibinfo {author} {\bibfnamefont {M.}~\bibnamefont
  {Yi}}, \bibinfo {author} {\bibfnamefont {Z.~K.}\ \bibnamefont {Liu}},
  \bibinfo {author} {\bibfnamefont {M.}~\bibnamefont {Hashimoto}}, \bibinfo
  {author} {\bibfnamefont {Y.}~\bibnamefont {Zhang}}, \bibinfo {author}
  {\bibfnamefont {D.~H.}\ \bibnamefont {Lu}}, \bibinfo {author} {\bibfnamefont
  {T.~P.}\ \bibnamefont {Devereaux}}, \bibinfo {author} {\bibfnamefont {D.-H.}\
  \bibnamefont {Lee}}, \ and\ \bibinfo {author} {\bibfnamefont {Z.-X.}\
  \bibnamefont {Shen}},\ }\href {http://dx.doi.org/10.1038/nature13894}
  {\bibfield  {journal} {\bibinfo  {journal} {Nature}\ }\textbf {\bibinfo
  {volume} {515}},\ \bibinfo {pages} {245} (\bibinfo {year}
  {2014})}\BibitemShut {NoStop}%
\bibitem [{\citenamefont {Ge}\ \emph {et~al.}(2014)\citenamefont {Ge},
  \citenamefont {Liu}, \citenamefont {Liu}, \citenamefont {Gao}, \citenamefont
  {Qian}, \citenamefont {Xue}, \citenamefont {Liu},\ and\ \citenamefont
  {Jia}}]{GeJF2014NatMat}%
  \BibitemOpen
  \bibfield  {author} {\bibinfo {author} {\bibfnamefont {J.-F.}\ \bibnamefont
  {Ge}}, \bibinfo {author} {\bibfnamefont {Z.-L.}\ \bibnamefont {Liu}},
  \bibinfo {author} {\bibfnamefont {C.}~\bibnamefont {Liu}}, \bibinfo {author}
  {\bibfnamefont {C.-L.}\ \bibnamefont {Gao}}, \bibinfo {author} {\bibfnamefont
  {D.}~\bibnamefont {Qian}}, \bibinfo {author} {\bibfnamefont {Q.-K.}\
  \bibnamefont {Xue}}, \bibinfo {author} {\bibfnamefont {Y.}~\bibnamefont
  {Liu}}, \ and\ \bibinfo {author} {\bibfnamefont {J.-F.}\ \bibnamefont
  {Jia}},\ }\href {http://dx.doi.org/10.1038/nmat4153} {\bibfield  {journal}
  {\bibinfo  {journal} {Nature Materials}\ }\textbf {\bibinfo {volume} {14}},\
  \bibinfo {pages} {285} (\bibinfo {year} {2014})}\BibitemShut {NoStop}%
\bibitem [{\citenamefont {Fang}\ \emph {et~al.}(2011)\citenamefont {Fang},
  \citenamefont {Wu}, \citenamefont {Thomale}, \citenamefont {Bernevig},\ and\
  \citenamefont {Hu}}]{FangC2011PRX}%
  \BibitemOpen
  \bibfield  {author} {\bibinfo {author} {\bibfnamefont {C.}~\bibnamefont
  {Fang}}, \bibinfo {author} {\bibfnamefont {Y.-L.}\ \bibnamefont {Wu}},
  \bibinfo {author} {\bibfnamefont {R.}~\bibnamefont {Thomale}}, \bibinfo
  {author} {\bibfnamefont {B.~A.}\ \bibnamefont {Bernevig}}, \ and\ \bibinfo
  {author} {\bibfnamefont {J.}~\bibnamefont {Hu}},\ }\href {\doibase
  10.1103/PhysRevX.1.011009} {\bibfield  {journal} {\bibinfo  {journal} {Phys.
  Rev. X}\ }\textbf {\bibinfo {volume} {1}},\ \bibinfo {pages} {011009}
  (\bibinfo {year} {2011})}\BibitemShut {NoStop}%
\bibitem [{\citenamefont {Zhou}\ \emph {et~al.}(2011)\citenamefont {Zhou},
  \citenamefont {Xu}, \citenamefont {Zhang},\ and\ \citenamefont
  {Chen}}]{ZhouY2011EPL}%
  \BibitemOpen
  \bibfield  {author} {\bibinfo {author} {\bibfnamefont {Y.}~\bibnamefont
  {Zhou}}, \bibinfo {author} {\bibfnamefont {D.-H.}\ \bibnamefont {Xu}},
  \bibinfo {author} {\bibfnamefont {F.-C.}\ \bibnamefont {Zhang}}, \ and\
  \bibinfo {author} {\bibfnamefont {W.-Q.}\ \bibnamefont {Chen}},\ }\href
  {http://stacks.iop.org/0295-5075/95/i=1/a=17003} {\bibfield  {journal}
  {\bibinfo  {journal} {EPL (Europhysics Letters)}\ }\textbf {\bibinfo {volume}
  {95}},\ \bibinfo {pages} {17003} (\bibinfo {year} {2011})}\BibitemShut
  {NoStop}%
\bibitem [{\citenamefont {Yang}\ \emph {et~al.}(2013)\citenamefont {Yang},
  \citenamefont {Wang},\ and\ \citenamefont {Lee}}]{YangF2013PRB}%
  \BibitemOpen
  \bibfield  {author} {\bibinfo {author} {\bibfnamefont {F.}~\bibnamefont
  {Yang}}, \bibinfo {author} {\bibfnamefont {F.}~\bibnamefont {Wang}}, \ and\
  \bibinfo {author} {\bibfnamefont {D.-H.}\ \bibnamefont {Lee}},\ }\href
  {\doibase 10.1103/PhysRevB.88.100504} {\bibfield  {journal} {\bibinfo
  {journal} {Phys. Rev. B}\ }\textbf {\bibinfo {volume} {88}},\ \bibinfo
  {pages} {100504} (\bibinfo {year} {2013})}\BibitemShut {NoStop}%
\bibitem [{\citenamefont {Maier}\ \emph {et~al.}(2011)\citenamefont {Maier},
  \citenamefont {Graser}, \citenamefont {Hirschfeld},\ and\ \citenamefont
  {Scalapino}}]{MaierTA2011PRB}%
  \BibitemOpen
  \bibfield  {author} {\bibinfo {author} {\bibfnamefont {T.~A.}\ \bibnamefont
  {Maier}}, \bibinfo {author} {\bibfnamefont {S.}~\bibnamefont {Graser}},
  \bibinfo {author} {\bibfnamefont {P.~J.}\ \bibnamefont {Hirschfeld}}, \ and\
  \bibinfo {author} {\bibfnamefont {D.~J.}\ \bibnamefont {Scalapino}},\ }\href
  {\doibase 10.1103/PhysRevB.83.100515} {\bibfield  {journal} {\bibinfo
  {journal} {Phys. Rev. B}\ }\textbf {\bibinfo {volume} {83}},\ \bibinfo
  {pages} {100515} (\bibinfo {year} {2011})}\BibitemShut {NoStop}%
\bibitem [{\citenamefont {Wang}\ \emph {et~al.}(2011)\citenamefont {Wang},
  \citenamefont {Yang}, \citenamefont {Gao}, \citenamefont {Lu}, \citenamefont
  {Xiang},\ and\ \citenamefont {Lee}}]{WangF2011EPL}%
  \BibitemOpen
  \bibfield  {author} {\bibinfo {author} {\bibfnamefont {F.}~\bibnamefont
  {Wang}}, \bibinfo {author} {\bibfnamefont {F.}~\bibnamefont {Yang}}, \bibinfo
  {author} {\bibfnamefont {M.}~\bibnamefont {Gao}}, \bibinfo {author}
  {\bibfnamefont {Z.-Y.}\ \bibnamefont {Lu}}, \bibinfo {author} {\bibfnamefont
  {T.}~\bibnamefont {Xiang}}, \ and\ \bibinfo {author} {\bibfnamefont {D.-H.}\
  \bibnamefont {Lee}},\ }\href {http://stacks.iop.org/0295-5075/93/i=5/a=57003}
  {\bibfield  {journal} {\bibinfo  {journal} {EPL (Europhysics Letters)}\
  }\textbf {\bibinfo {volume} {93}},\ \bibinfo {pages} {57003} (\bibinfo {year}
  {2011})}\BibitemShut {NoStop}%
\bibitem [{\citenamefont {Mazin}(2011)}]{MazinII2011PRB}%
  \BibitemOpen
  \bibfield  {author} {\bibinfo {author} {\bibfnamefont {I.~I.}\ \bibnamefont
  {Mazin}},\ }\href {\doibase 10.1103/PhysRevB.84.024529} {\bibfield  {journal}
  {\bibinfo  {journal} {Phys. Rev. B}\ }\textbf {\bibinfo {volume} {84}},\
  \bibinfo {pages} {024529} (\bibinfo {year} {2011})}\BibitemShut {NoStop}%
\bibitem [{\citenamefont {Fan}\ \emph {et~al.}(2015)\citenamefont {Fan},
  \citenamefont {Zhang}, \citenamefont {Liu}, \citenamefont {Yan},
  \citenamefont {Ren}, \citenamefont {Peng}, \citenamefont {Xu}, \citenamefont
  {Xie}, \citenamefont {Hu}, \citenamefont {Zhang},\ and\ \citenamefont
  {Feng}}]{FanQ2015NatPhys}%
  \BibitemOpen
  \bibfield  {author} {\bibinfo {author} {\bibfnamefont {Q.}~\bibnamefont
  {Fan}}, \bibinfo {author} {\bibfnamefont {W.~H.}\ \bibnamefont {Zhang}},
  \bibinfo {author} {\bibfnamefont {X.}~\bibnamefont {Liu}}, \bibinfo {author}
  {\bibfnamefont {Y.~J.}\ \bibnamefont {Yan}}, \bibinfo {author} {\bibfnamefont
  {M.~Q.}\ \bibnamefont {Ren}}, \bibinfo {author} {\bibfnamefont
  {R.}~\bibnamefont {Peng}}, \bibinfo {author} {\bibfnamefont {H.~C.}\
  \bibnamefont {Xu}}, \bibinfo {author} {\bibfnamefont {B.~P.}\ \bibnamefont
  {Xie}}, \bibinfo {author} {\bibfnamefont {J.~P.}\ \bibnamefont {Hu}},
  \bibinfo {author} {\bibfnamefont {T.}~\bibnamefont {Zhang}}, \ and\ \bibinfo
  {author} {\bibfnamefont {D.~L.}\ \bibnamefont {Feng}},\ }\href
  {http://dx.doi.org/10.1038/nphys3450} {\bibfield  {journal} {\bibinfo
  {journal} {Nature Physics}\ }\textbf {\bibinfo {volume} {11}},\ \bibinfo
  {pages} {946} (\bibinfo {year} {2015})}\BibitemShut {NoStop}%
\bibitem [{\citenamefont {Kang}\ and\ \citenamefont
  {Fernandes}(2016)}]{KangJ2016PRL}%
  \BibitemOpen
  \bibfield  {author} {\bibinfo {author} {\bibfnamefont {J.}~\bibnamefont
  {Kang}}\ and\ \bibinfo {author} {\bibfnamefont {R.~M.}\ \bibnamefont
  {Fernandes}},\ }\href {\doibase 10.1103/PhysRevLett.117.217003} {\bibfield
  {journal} {\bibinfo  {journal} {Phys. Rev. Lett.}\ }\textbf {\bibinfo
  {volume} {117}},\ \bibinfo {pages} {217003} (\bibinfo {year}
  {2016})}\BibitemShut {NoStop}%
\bibitem [{\citenamefont {Zhang}\ \emph {et~al.}(2016)\citenamefont {Zhang},
  \citenamefont {Lee}, \citenamefont {Moore}, \citenamefont {Li}, \citenamefont
  {Yi}, \citenamefont {Hashimoto}, \citenamefont {Lu}, \citenamefont
  {Devereaux}, \citenamefont {Lee},\ and\ \citenamefont
  {Shen}}]{ZhangY2016PRL}%
  \BibitemOpen
  \bibfield  {author} {\bibinfo {author} {\bibfnamefont {Y.}~\bibnamefont
  {Zhang}}, \bibinfo {author} {\bibfnamefont {J.~J.}\ \bibnamefont {Lee}},
  \bibinfo {author} {\bibfnamefont {R.~G.}\ \bibnamefont {Moore}}, \bibinfo
  {author} {\bibfnamefont {W.}~\bibnamefont {Li}}, \bibinfo {author}
  {\bibfnamefont {M.}~\bibnamefont {Yi}}, \bibinfo {author} {\bibfnamefont
  {M.}~\bibnamefont {Hashimoto}}, \bibinfo {author} {\bibfnamefont {D.~H.}\
  \bibnamefont {Lu}}, \bibinfo {author} {\bibfnamefont {T.~P.}\ \bibnamefont
  {Devereaux}}, \bibinfo {author} {\bibfnamefont {D.-H.}\ \bibnamefont {Lee}},
  \ and\ \bibinfo {author} {\bibfnamefont {Z.-X.}\ \bibnamefont {Shen}},\
  }\href {\doibase 10.1103/PhysRevLett.117.117001} {\bibfield  {journal}
  {\bibinfo  {journal} {Phys. Rev. Lett.}\ }\textbf {\bibinfo {volume} {117}},\
  \bibinfo {pages} {117001} (\bibinfo {year} {2016})}\BibitemShut {NoStop}%
\bibitem [{\citenamefont {Agterberg}\ \emph {et~al.}(2017)\citenamefont
  {Agterberg}, \citenamefont {Shishidou}, \citenamefont {O'Halloran},
  \citenamefont {Brydon},\ and\ \citenamefont {Weinert}}]{AgterbergDF2017PRL}%
  \BibitemOpen
  \bibfield  {author} {\bibinfo {author} {\bibfnamefont {D.~F.}\ \bibnamefont
  {Agterberg}}, \bibinfo {author} {\bibfnamefont {T.}~\bibnamefont
  {Shishidou}}, \bibinfo {author} {\bibfnamefont {J.}~\bibnamefont
  {O'Halloran}}, \bibinfo {author} {\bibfnamefont {P.~M.~R.}\ \bibnamefont
  {Brydon}}, \ and\ \bibinfo {author} {\bibfnamefont {M.}~\bibnamefont
  {Weinert}},\ }\href {\doibase 10.1103/PhysRevLett.119.267001} {\bibfield
  {journal} {\bibinfo  {journal} {Phys. Rev. Lett.}\ }\textbf {\bibinfo
  {volume} {119}},\ \bibinfo {pages} {267001} (\bibinfo {year}
  {2017})}\BibitemShut {NoStop}%
\bibitem [{\citenamefont {Korshunov}\ and\ \citenamefont
  {Eremin}(2008)}]{Korshunov2008PRB}%
  \BibitemOpen
  \bibfield  {author} {\bibinfo {author} {\bibfnamefont {M.~M.}\ \bibnamefont
  {Korshunov}}\ and\ \bibinfo {author} {\bibfnamefont {I.}~\bibnamefont
  {Eremin}},\ }\href {\doibase 10.1103/PhysRevB.78.140509} {\bibfield
  {journal} {\bibinfo  {journal} {Phys. Rev. B}\ }\textbf {\bibinfo {volume}
  {78}},\ \bibinfo {pages} {140509} (\bibinfo {year} {2008})}\BibitemShut
  {NoStop}%
\bibitem [{\citenamefont {Cao}\ \emph {et~al.}(2014)\citenamefont {Cao},
  \citenamefont {Tan}, \citenamefont {Xiang}, \citenamefont {Feng},\ and\
  \citenamefont {Gong}}]{CaoHY2014PRB}%
  \BibitemOpen
  \bibfield  {author} {\bibinfo {author} {\bibfnamefont {H.-Y.}\ \bibnamefont
  {Cao}}, \bibinfo {author} {\bibfnamefont {S.}~\bibnamefont {Tan}}, \bibinfo
  {author} {\bibfnamefont {H.}~\bibnamefont {Xiang}}, \bibinfo {author}
  {\bibfnamefont {D.~L.}\ \bibnamefont {Feng}}, \ and\ \bibinfo {author}
  {\bibfnamefont {X.-G.}\ \bibnamefont {Gong}},\ }\href {\doibase
  10.1103/PhysRevB.89.014501} {\bibfield  {journal} {\bibinfo  {journal} {Phys.
  Rev. B}\ }\textbf {\bibinfo {volume} {89}},\ \bibinfo {pages} {014501}
  (\bibinfo {year} {2014})}\BibitemShut {NoStop}%
\bibitem [{\citenamefont {Cassanello}\ and\ \citenamefont
  {Fradkin}(1996)}]{Cassanello1996PRB}%
  \BibitemOpen
  \bibfield  {author} {\bibinfo {author} {\bibfnamefont {C.~R.}\ \bibnamefont
  {Cassanello}}\ and\ \bibinfo {author} {\bibfnamefont {E.}~\bibnamefont
  {Fradkin}},\ }\href {\doibase 10.1103/PhysRevB.53.15079} {\bibfield
  {journal} {\bibinfo  {journal} {Phys. Rev. B}\ }\textbf {\bibinfo {volume}
  {53}},\ \bibinfo {pages} {15079} (\bibinfo {year} {1996})}\BibitemShut
  {NoStop}%
\bibitem [{\citenamefont {Cassanello}\ and\ \citenamefont
  {Fradkin}(1997)}]{Cassanello1997PRB}%
  \BibitemOpen
  \bibfield  {author} {\bibinfo {author} {\bibfnamefont {C.~R.}\ \bibnamefont
  {Cassanello}}\ and\ \bibinfo {author} {\bibfnamefont {E.}~\bibnamefont
  {Fradkin}},\ }\href {\doibase 10.1103/PhysRevB.56.11246} {\bibfield
  {journal} {\bibinfo  {journal} {Phys. Rev. B}\ }\textbf {\bibinfo {volume}
  {56}},\ \bibinfo {pages} {11246} (\bibinfo {year} {1997})}\BibitemShut
  {NoStop}%
\bibitem [{\citenamefont {Withoff}\ and\ \citenamefont
  {Fradkin}(1990)}]{Withoff1990PRL}%
  \BibitemOpen
  \bibfield  {author} {\bibinfo {author} {\bibfnamefont {D.}~\bibnamefont
  {Withoff}}\ and\ \bibinfo {author} {\bibfnamefont {E.}~\bibnamefont
  {Fradkin}},\ }\href {\doibase 10.1103/PhysRevLett.64.1835} {\bibfield
  {journal} {\bibinfo  {journal} {Phys. Rev. Lett.}\ }\textbf {\bibinfo
  {volume} {64}},\ \bibinfo {pages} {1835} (\bibinfo {year}
  {1990})}\BibitemShut {NoStop}%
\bibitem [{\citenamefont {Gonzalez-Buxton}\ and\ \citenamefont
  {Ingersent}(1998)}]{GonzalezBuxton1998PRB}%
  \BibitemOpen
  \bibfield  {author} {\bibinfo {author} {\bibfnamefont {C.}~\bibnamefont
  {Gonzalez-Buxton}}\ and\ \bibinfo {author} {\bibfnamefont {K.}~\bibnamefont
  {Ingersent}},\ }\href {\doibase 10.1103/PhysRevB.57.14254} {\bibfield
  {journal} {\bibinfo  {journal} {Phys. Rev. B}\ }\textbf {\bibinfo {volume}
  {57}},\ \bibinfo {pages} {14254} (\bibinfo {year} {1998})}\BibitemShut
  {NoStop}%
\bibitem [{\citenamefont {Zhuang}\ \emph {et~al.}(2009)\citenamefont {Zhuang},
  \citenamefont {feng Sun},\ and\ \citenamefont {Xie}}]{ZhuangHB2009EPL}%
  \BibitemOpen
  \bibfield  {author} {\bibinfo {author} {\bibfnamefont {H.-B.}\ \bibnamefont
  {Zhuang}}, \bibinfo {author} {\bibfnamefont {Q.}~\bibnamefont {feng Sun}}, \
  and\ \bibinfo {author} {\bibfnamefont {X.~C.}\ \bibnamefont {Xie}},\ }\href
  {http://stacks.iop.org/0295-5075/86/i=5/a=58004} {\bibfield  {journal}
  {\bibinfo  {journal} {EPL (Europhysics Letters)}\ }\textbf {\bibinfo {volume}
  {86}},\ \bibinfo {pages} {58004} (\bibinfo {year} {2009})}\BibitemShut
  {NoStop}%
\bibitem [{\citenamefont {Uchoa}\ \emph {et~al.}(2011)\citenamefont {Uchoa},
  \citenamefont {Rappoport},\ and\ \citenamefont
  {Castro~Neto}}]{UchoaB2012PRL}%
  \BibitemOpen
  \bibfield  {author} {\bibinfo {author} {\bibfnamefont {B.}~\bibnamefont
  {Uchoa}}, \bibinfo {author} {\bibfnamefont {T.~G.}\ \bibnamefont
  {Rappoport}}, \ and\ \bibinfo {author} {\bibfnamefont {A.~H.}\ \bibnamefont
  {Castro~Neto}},\ }\href {\doibase 10.1103/PhysRevLett.106.016801} {\bibfield
  {journal} {\bibinfo  {journal} {Phys. Rev. Lett.}\ }\textbf {\bibinfo
  {volume} {106}},\ \bibinfo {pages} {016801} (\bibinfo {year}
  {2011})}\BibitemShut {NoStop}%
\end{thebibliography}%

\end{document}